\newcommand{\pcmq}{\mbox{cm$^{-2}$}}
\newcommand{\psec}{\mbox{s$^{-1}$}}

\newcommand{\funit}{\mbox{ph~\pcmq~\psec}}
\newcommand{\iunit}{\mbox{\funit~sr$^{-1}$}}
\newcommand{\gam}{$\gamma$}
\newcommand{\degree}{\ensuremath{^\circ}}
\newcommand{\fermiLAT}{\textit{Fermi}-LAT}
\newcommand{\Nmsp}{36}
\newcommand{\Nuna}{{66}}

\newcommand{\stexp}{\mbox{\texttt{ST1}}}
\newcommand{\stgau}{\mbox{\texttt{ST2}}}
\newcommand{\fgexp}{\mbox{\texttt{FG1}}}
\newcommand{\fggau}{\mbox{\texttt{FG2}}}
\newcommand{\stmsp}{\mbox{\stexp}}
\newcommand{\stuna}{\mbox{\texttt{UNA}}}
\newcommand{\stst}{\mbox{\texttt{SIMU36}}}
\newcommand{\stsh}{\mbox{\texttt{SIMU66}}}
\newcommand{\Fcr}{\mbox{$S_{|b| \le 10\degree,\ |l|\le 30\degree}$}}
\newcommand{\Ipsr}{\mbox{$S_{|b| \ge 40\degree}$}}
\newcommand{\Nexp}{\mbox{$N_{\rm exp}$}}
\newcommand{\sensFact}{\mbox{1.4}}

\documentclass[structabstract]{aa}  

\usepackage{txfonts} 
\usepackage{graphicx} 
\usepackage[left]{lineno} 
\usepackage{natbib}
\usepackage[tight]{subfigure} 
\usepackage{multirow}
\bibpunct{(}{)}{;}{a}{}{,} 
\title{Constraining the Galactic millisecond pulsar population using Fermi Large Area Telescope}
\titlerunning{Constraining Galactic MSP population using \fermiLAT}

\author{Tristan Gr\'egoire\inst{\ref{inst1}}
\and J\"urgen Kn\"odlseder\inst{\ref{inst1}}}
\authorrunning{Tristan Gr\'egoire \& J\"urgen Kn\"odlseder}

\institute{Institut de Recherche en Astrophysique et Plan\'etologie, UPS/CNRS, 31028 Toulouse Cedex 4, France\label{inst1}\\
\email{tristan.gregoire@irap.omp.eu}}

\date{Received March 15, 2013; accepted March 15, 2013}

\abstract
{The Fermi Large Area Telescope (\fermiLAT) has recently revealed a large population 
of gamma-ray emitting millisecond pulsars (MSPs) in our Galaxy.
}
{We aim to infer the properties of the Galactic population of gamma-ray emitting MSPs
from the samples detected by the \fermiLAT.
}
{We developed a Monte Carlo model to predict the spatial and gamma-ray luminosity distribution 
of the Galactic MSP population.
Based on the estimated detection sensitivity of \fermiLAT, we split the model population into
{\em detectable \em} and {\em undetectable\em} samples of MSPs.
Using a maximum likelihood method, we compared the {\em detectable\em} sample to a set of
\Nmsp\ MSPs detected by \fermiLAT, and we derived the parameters of the spatial distribution
and the total number of gamma-ray emitting MSPs in the Galaxy.
The corresponding {\em undetectable\em} sample provided us with an estimate for the expected
diffuse emission from unresolved MSPs in the Milky Way.
We also applied our method to an extended sample of \Nuna\ MSPs that combines firmly
detected MSPs and \gam-ray sources that show characteristics reminiscent of MSPs.
}
{
Using the sample of \Nmsp\ MSPs detected by \fermiLAT, our
analysis suggests the existence of $9\,000-11\,000$ \gam-ray emitting MSPs in the
Galaxy.
The maximum likelihood analysis suggests 
an exponential radial scale length of $\sim 4$~kpc and 
an exponential vertical scale height of $\sim 1$~kpc for the underlying 
MSP population.
The unresolved population of Galactic \gam-ray emitting MSPs is predicted to
contribute a flux of $\sim 2 \times 10^{-6}$~\iunit\ to the Galactic diffuse emission 
observed from the central radian above 100 MeV.
This value corresponds to $\sim$~1\% of the total observed \gam-ray flux from that region.
For latitudes $|b|\ge 40\degree$ the expected average intensity amounts to
$\sim 2 \times 10^{-8}$~\iunit\ above 100 MeV, which corresponds to $0.2\%$ of the 
high-latitude background intensity.
Using the extended sample increases the estimated number of \gam-ray emitting MSPs
in the Galaxy to $\sim22\,000$ and slightly reduces the scale parameters of the spatial
distribution.
The results are robust with respect to systematic uncertainties in the estimated \fermiLAT\ 
detection sensitivity.
}
{For the first time our analysis provides \gam-ray based constraints on the Galactic population
of MSPs.
The radial scale length and vertical scale height of the population is consistent with estimates
based on radio data.
Our analysis suggests that MSPs do not provide any significant contribution to the isotropic 
diffuse \gam-ray background emission.
}


\begin{document}

\maketitle

\section{Introduction}

Pulsars are highly magnetized and rapidly spinning neutron stars, obeying a beam of radiation
that periodically intersects the Earth.
Pulsars were first discovered in the radio band by \citet{Hewish1968}, but they are observed
today throughout the electromagnetic spectrum, which covers visible light, radio waves, X-rays and 
gamma-rays.
Several classes of pulsars are distinguishable, characterized by their spin period and period
derivative. The period derivative relates to the spin down power, characteristic pulsar age, and surface 
magnetic field.
Ironically, the most rapidly spinning pulsars that reach spin periods of less than a few tens
of milliseconds are also the oldest pulsars.
These so-called millisecond pulsars (MSPs) are commonly interpreted as the result
of a mass-transfer period in a close binary system, where the spin of an
old neutron star has been drastically increased by the transfer of angular momentum
\citep{Alpar1982}.

Radio observations have so far unveiled a population of 137 Galactic field MSPs,
while the Galactic population of MSPs has been estimated to $\sim30\,000$
\citep{Cordes1997, Lyne1998, Lorimer2005}.
Improving our knowledge of the Galactic MSP population provides important clues
to understanding the MSP progenity.
In this work, we try to constrain the Galactic MSP population using \gam-ray observations of
MSPs.
Before the launch of the Fermi satellite, only circumstantial evidence of pulsed \gam-rays from 
one MSP had been found \citep{Kuiper2000}.
Now, the Large Area Telescope (LAT) aboard Fermi has been
established them as a new and important population of Galactic \gam-ray sources
\citep{Abdo2009}.
So far, more than 40 MSPs have been detected by \fermiLAT, which corresponds to 
$\sim30\%$
of the Galactic field MSP population.
Moreover, the search for radio pulsations towards the direction of sitll unidentified \gam-ray
sources that obey spectral characteristics of typical \gam-ray pulsars has turned out to be extremely
efficient to unveil yet unknown MSPs \citep[e.g.][]{Cognard2011, Ransom2011, Kerr2012}.

We make use of this new sample of Galactic field \gam-ray MSPs to constrain the
total number and spatial distribution of \gam-ray emitting MSPs within our Galaxy.
In section \ref{sec:data}, we describe the sample of \gam-ray emitting MSPs used in this
study.
In section \ref{sec:model}, we introduce a population model of Galactic MSPs that we
use to constrain the population parameters in section \ref{sec:const}.
The results are discussed in section \ref{sec:dis}, and we conclude in section \ref{sec:conclusion}.

\section{MSP sample used in this study\label{sec:data}}

\begin{table}
  \label{tab:msp}
  \caption{List of \Nmsp\ MSPs seen by the \fermiLAT\ and used in this study.}
  \centering
  \begin{tabular}{lrrc}
    \hline\hline
     1FGL Name & $l$ (\degree) & $b$ (\degree) & References \\
    \hline
     J0023.5+0930 & 111.523 & -52.743 & A \\
     J0030.4+0451 & 113.142 & -57.611 & 1, 2, 3 \\
     J0034.3$-$0534 & 111.493 & -68.069 & 4 \\
     J0101.0$-$6423 & 301.219 & -52.700 & 10 \\
     J0103.1+4840 & 124.933 & -14.155 & A \\
     J0218.1+4232 & 139.509 & -17.527 & 2, 3 \\
     J0340.4+4130 & 153.794 & -11.022 & P1 \\
     J0437.2$-$4715 & 253.395 & -41.964 & 2, 3 \\
     J0610.7$-$2059 & 227.786 & -18.071 & P3 \\
     J0613.7$-$0200 & 210.413 & -9.3047 & 2, 3 \\
     J0614.1$-$3328 & 240.482 & -21.819 & 5 \\
     J0751.1+1807 & 202.730 &  21.086 & 2, 3 \\
     J1024.6$-$0718 & 251.702 &  40.524 & P3 \\
     J1124.4$-$3654 & 284.189 &  22.772 & P2 \\
     J1231.1$-$1410 & 295.529 &  48.406 & 5 \\
     J1446.8$-$4702 & 322.527 &  11.394 & 8 \\
     J1514.1$-$4945 & 325.229 &   6.832 & A \\
     J1600.7$-$3055 & 344.045 &  16.452 & P3 \\
     J1614.0$-$2230 & 352.541 &  20.301 & 2, 3 \\
     J1658.8$-$5317 & 334.977 &  -6.577 & 10 \\
     J1713.9+0750 &  28.820 &  25.210 & P3 \\
     J1744.4$-$1134 &  14.794 &   9.179 & 2, 3 \\
     J1747.4$-$4035 & 350.195 &  -6.338 & 10 \\
     J1810.3+1741 &  44.570 &  16.840 & A \\
     J1858.1$-$2218 &  13.537 & -11.373 & P4 \\
     J1902.0$-$5110 & 345.579 & -22.405 & 10 \\
     J1938.2+2125 &  57.207 &  -0.092 & 9 \\
     J1959.6+2047 &  59.193 &  -4.703 & 9 \\
     J2017.3+0603 &  48.623 & -16.020 & 6 \\
     J2043.2+1709 &  61.887 & -15.317 & 11 \\
     J2047.6+1055 &  57.159 & -19.750 & P4 \\
     J2124.7$-$3358 &  10.926 & -45.438 & 2, 3 \\
     J2214.8+3002 &  86.909 & -21.663 & 5 \\
     J2216.1+5139 &  99.979 &  -4.154 & A \\
     J2241.9$-$5236 & 337.420 & -54.950 & 7 \\
     J2302.8+4443 & 103.415 & -13.984 & 6 \\
     \hline
  \end{tabular}
  \tablefoot{$l$ and $b$ are Galactic longitude and latitude in degrees, respectively.}
  \tablebib{We distinguish conference proceeding by adding P to the reference. 
  (A)~\citet{ATNFpaper};
  (1)~\citet{Abdo2009c}; (2)~\citet{Abdo2009}; (3)~\citet{Abdo2010a}; (4)~\citet{Abdo2010c};
  (5)~\citet{Ransom2011}; (6)~\citet{Cognard2011}; (7)~\citet{Keith2011}; (8)~\citet{Keith2012};
  (9)~\citet{Guillemot2012a}; (10)~\citet{Kerr2012}; (11)~\citet{Guillemot2012};
  (P1)~\citet{Guillemot2010}; (P2)~\citet{Hessels2010}; (P3)~\citet{Parent2010}; (P4)~\citet{Ray2011a}
  }
\end{table}


The sample of \gam-ray MSPs used in this work is based on the first catalogue of \fermiLAT\
sources \citep[1FGL;][]{Abdo2010a}.
Among the 1451 sources found in the 1FGL catalogue, 821 had been associated to known
sources at other wavelengths, while 630 sources remained unassociated.
By the time the 1FGL catalogue was published, 9 sources in the catalogue
were identified as MSPs through their observed \gam-ray pulsations.
All \gam-ray MSPs detected in 1FGL show similar spectral shapes, exhibiting hard
power laws with exponential cut-offs near a few GeV.
A non-negligible fraction of the unassociated sources exhibited similar spectral features,
and dedicated searches for pulsations using radio searches of the corresponding 1FGL error 
boxes have led to the discovery of 27 still unknown Galactic field MSPs.

This result established a new strategy of discovering Galactic MSPs:
the \fermiLAT\ telescope indicates potential locations of Galactic MSPs that are then
investigated using radio observations to reveal the pulsar's pulsations.
Although these studies are still limited by the sensitivity of the radio telescopes to unveil
the radio pulsations, they are probably less biased than the pure radio surveys, because
\fermiLAT\ scans the entire sky rather homogeneously.
It can thus be expected that the \fermiLAT\ sample of Galactic MSPs is mainly limited
by the actual sensitivity of the \gam-ray observations.
In total, \Nmsp\ \gam-ray MSPs have been firmly identified in the 1FGL catalogue.
We use this sample in this paper and summarize their 1FGL names,
together with the source positions and relevant references in Table \ref{tab:msp}. 

Nevertheless, we also recognize the possibility that this sample is not totally free of any radio
bias, as the identification of the MSPs still relies on the detection of the pulsar at
radio frequencies.
We thus also consider an extended sample of \Nuna\ objects in this work that is comprised
of identified \gam-ray MSPs and unassociated 1FGL sources, which are located outside the Galactic plane
with spectral and temporal characteristics reminiscent of Galactic MSPs.
As probably not all sources in this sample are indeed MSPs, we may consider this sample
as an upper limit to the true sample of Galactic \gam-ray emitting MSPs detectable by
\fermiLAT\ during the first 11 months of observation that served as the basis for the 1FGL
source catalogue.

\section{MSP population model\label{sec:model}}

\subsection{General approach}

With characteristic ages of the order of $10^9$ yr, MSPs are members of the old stellar
population, which can be described by a relatively smooth Galactic density distribution 
with a vertical scale height of several $100$ pc \citep{Robin2003}.
Dynamic modelling of the MSP evolution in the Galactic gravitational potential suggests
radial and vertical scale heights of
$R_0 \approx 4.2$~kpc and
$z_0 \approx 0.5$~kpc, assuming exponential spatial distributions of the form
\begin{equation}
\rho(R,z) \propto \exp{(-R/R_0)} \exp{(-|z|/z_0)}
\label{eq:expspatial}
\end{equation}
where $R$ is the distance from the Galactic centre and $z$ is the vertical scale height
measured from the Galactic plane
\citep{Story2007}.
Other authors model the radial distribution of MSPs using a Gaussian
density profile
\begin{equation}
\rho(R,z) \propto \exp{(-R^2/2\sigma_r^2)} \exp{(-|z|/z_0)}
\label{eq:gaussspatial}
\end{equation}
with $\sigma_r \approx 5$~kpc and $z_0 \approx 1$~kpc \citep{Faucher-Giguere2010}.
We test both equations in the present work, but we already note in advance that our
results are fairly insensitive to the selected radial law.
We do not attempt to model a possible contribution from Galactic bulge MSPs, because \fermiLAT\ 
observations currently probe only the nearby MSP population; thus, the bulge contribution
essentially remains unconstrained.

\begin{figure*}
  \centering
  \includegraphics[width=\textwidth]{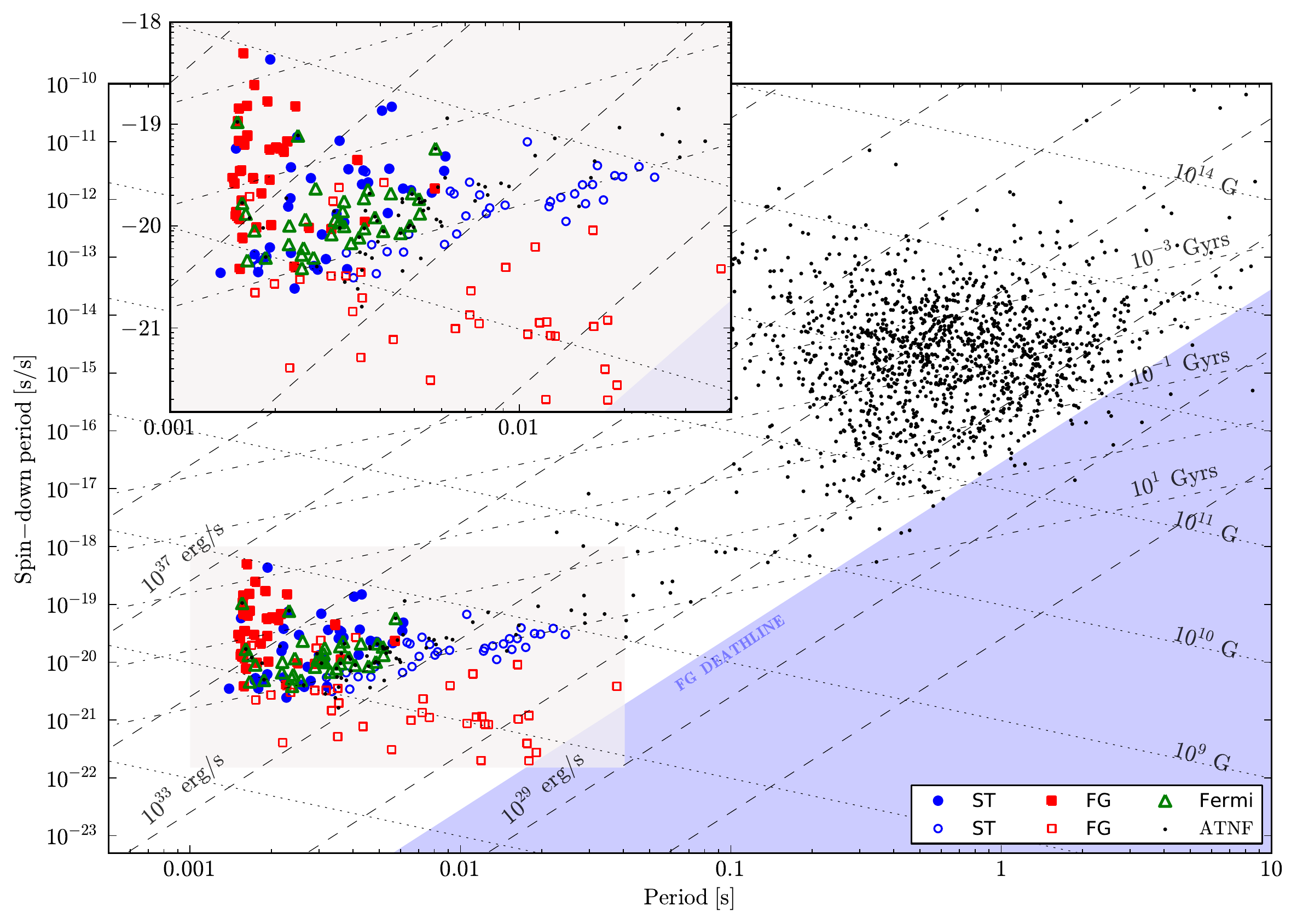}
  \caption{Simulated $P-\dot P$ distribution of 
  the {\em detectable} sample (filled symbols)
  and an according {\em undetectable} sample (open symbols)
  using the approaches of 
  \citet{Faucher-Giguere2010} (red squares) and of \citet{Story2007} (blue cirles).
  We use [$\sigma_{\rm r}$ = 5, $z_{\rm 0}$ = 1] kpc and [$R_{\rm 0}$ = 4.2, $z_{\rm 0}$ = 0.5]
  kpc respectively for \citeauthor{Faucher-Giguere2010} and \citeauthor{Story2007} approaches.
  For comparison we show the observed distribution of \fermiLAT\ MSPs as green triangles.
  The $P-\dot P$ distribution of all pulsars from the ATNF catalogue
  \citep[\texttt{http://www.atnf.csiro.au/research/pulsar/psrcat}]{ATNFpaper} is shown as black dots.
  We also show lines of constant spin-down power $\dot E$ (dashed), constant characteristic
  age (dashed-dotted), and constant magnetic field (dotted).
  The area excluded by the deathline implemented in \citeauthor{Faucher-Giguere2010}'s
  model is indicated by the blue filled area.
  The inset presents a zoom into the MSP region of the $P-\dot P$ diagram.}
  \label{fig:PPdot}
\end{figure*}

The first step in our MSP population modelling procedure is that we draw random locations
of pulsars from the density laws Eqs.~(\ref{eq:expspatial}) or (\ref{eq:gaussspatial}).
By definition, the spatial distributions are axisymmetric with respect to the Galactic
centre, and for randomly drawn pairs of $R$ and $z$, we randomly assign an azimuth
angle by drawing from a uniform distribution between $0$ and $2\pi$.
We then transform these Galactocentric coordinates into 
Galactic longitude $l$,
latitude $b$,
and distance $d$ from the Sun by assuming that the Sun is placed at $8.5$ kpc from
the Galactic centre and $20$ pc above the Galactic plane \citep{Reed2006}.

As a second step, we assign for each pulsar a period $P$ and a period derivative
$\dot P$ using either 
the approach proposed by \citet{Faucher-Giguere2010} (FG),
or the approach developed by \citet{Story2007} (ST; see section~\ref{sec:paramLg}).
We do not explicitly take orientation or light beam effects into account, but we 
assume that each MSP obeys a mean luminosity drawn from a luminosity distribution 
function that only depends on $P$ and $\dot P$ (see section~\ref{sec:luminosity}).

Finally, we then determine the \gam-ray
flux $F_{\gamma}$ on Earth from the assigned luminosity using
\begin{equation}
F_{\gamma} = \frac{L_{\gamma}}{\Delta \Omega d^2} \, .
\label{eq:flux}
\end{equation}
where we take $\Delta \Omega=4 \pi$.

This procedure provides us with a list of positions and fluxes for a population of MSPs that is
then used for further analysis (section~\ref{sec:const}).
We typically draw samples of $N_{\rm MC} = 10^7$ pulsars in the Galaxy to reduce the
impact of statistical fluctuations on our final results.
We then split the sample into two subsamples.
The first subsample contains all pulsars for which $F_{\gamma} > F_{\rm sensitivity}$, where 
$F_{\rm sensitivity}$ is the detection sensitivity of \fermiLAT\ at the specific sky position of the 
pulsar.
We call this subsample the {\em detectable MSP} population.
The second subsample contains all pulsars with fluxes below the sensitivity limit,
making up the subsample of {\em undetectable MSPs}.

We use the {\em detectable MSP} population as a reference for comparison to
the observed \gam-ray MSP population, to draw conclusions about the
spatial parameters $R_\mathrm{0}/\sigma_{\rm r}$ and $z_\mathrm{0}$ and the 
total number $N_{\rm MSP}$ of 
\gam-ray emitting MSPs in our Galaxy (cf.~section \ref{sec:const}).
The subsample of {\em undetectable MSPs} is used to study the Galactic diffuse
emission that will arise from the large population of pulsars falling below the
detection threshold (cf.~section~\ref{sec:dis}).

\subsection{MSP period and period derivative\label{sec:paramLg}}
\subsubsection{Faucher-Gigu\`ere \& Loeb (2010)\label{sec:fg}}
Our first method to assign $P$ and $\dot P$ to individual pulsars has been inspired from 
the work of \citet{Faucher-Giguere2010}.
We randomly draw $P$ from the empirically determined power law distribution
$N(P) \propto P^{-1}$
\citep{Cordes1997} under the constraint
$P \ge 1.5$ ms.
We also randomly draw a value for the magnetic field $B$ from the distribution
\begin{equation}
\label{eq:cafgB}
N(\log B) \propto \exp \left( -{\left( \log B - \langle \log B \rangle \right)}^{2} / 2\sigma_{\rm log B}^2 \right)
\end{equation}
with $\langle \log B \rangle = 8$ and $\sigma_{\rm log B} = 0.3$.
Using both quantities, we then use the conventional formula for magnetic dipol braking,
which is expressed as
$B=3.2 \times 10^{19} (P \dot P)^{1/2}$ G \citep{Lyne2000},
to derive $\dot P$ for each pulsar.
We then introduce a ``deathline" defined by 
$B/P^2 = 0.17 \times 10^{12} $~$ \mathrm{G \cdot s^{-2}}$
and only retain pulsars that lie above this deathline in the $P-\dot P$ diagram.
We note that the exponent of $N(P)$ and the spread of the magnetic field $\sigma_{\rm log B}$
that we have chosen differ slightly from the values adopted by \citet{Faucher-Giguere2010}
(they use $N(P) \propto P^{-2}$ and $\sigma_{\rm log B} = 0.2$), because we found that our values
produce a slightly better match to the observed $P-\dot P$ distribution.

Figure~\ref{fig:PPdot} shows the simulated $P-\dot P$ distribution of a sample of
\emph{detectable} (red filled squares) and \emph{undetectable} (red open squares) pulsars.
For the shake of comparison to the observed sample of \Nmsp\ MSPs, we limit here the total number of 
simulated pulsars to the first $N$ pulsars that comprise \Nmsp\ detectable MSPs.
Apparently, the approach of \citet{Faucher-Giguere2010} produces a non-negligible
population of undetectable MSPs with characteristics ages largely exceeding the age of the 
Universe.
Although these pulsars are classified as \emph{undetectable} by Fermi-LAT, they contribute
to some extent to the diffuse \gam-ray emission of the Galaxy, and they affect our estimate
of the total number of \gam-ray emitting MSPs that are present in the Milky Way.

\subsubsection{Story et al. (2007)\label{sec:st}}
Our second method to assign $P$ and $\dot P$ to individual pulsars has been based 
on the work of \citet{Story2007}.
As a first step, we draw a random magnetic field from within the interval
$[B_{\rm min}, B_{\rm max}]=[0.1,10] \times 10^{8}$~G 
using the distribution
$N(B_{8}) = \left(B_8 \ln \left( B_{\rm max} / B_{\rm min} \right) \right)^{-1}$,
where 
$B_8$ is the magnetic field in units of $10^8$ G.
 \citet{Story2007} use a larger range of $[1,10^4] \times 10^{8}$~G
in their simulations, but we found that setting the range to $[0.1,10] \times 10^{8}$~G
improves the match between the observed and predicted $P - \dot P$ distribution.
The initial spin period of MSPs is then determined using
$P_0 = 0.18 \times 10^{3\delta/7} B_8^{6/7}$,
where $\delta$ is a dithering parameter introduced to take variations in the
accretion rate of the MSP progenitors into account.
Similar to \citet{Story2007}, we draw $\delta$ from a ramp distribution that
increases by a factor of 4 between 0 and 2.8.

We then draw an age $\tau$ for each MSP using a uniform age distribution within
$[0,12]$ Gyr, and based on this age and assuming a constant dipole magnetic field,
we spin-down the initial spin period to the current spin-period using
$P = \left( 1.95 \times 10^{-23} \tau B_8^2 + P_0^2 \right)^{-1/2}$,

We then estimate the derivative period, using $\dot P = 9.77 \times 10^{-24} B_8^2 / P$.

Finally, we force the initial period $P_0$ to be greater than 1.3~ms and
we take electron-positron pair production into account that occurs for
strong fields and/or long spin periods \citep{Alpar1982} by requiring
$B_{12}P^{-2} \ge 1$, which is typical of a pair production model
\citep[no cut-off has been considered by][]{Story2007}.

The resulting $P - \dot P$ distribution is shown in Fig.~\ref{fig:PPdot}, using filled blue circles
for the \emph{detectable} and open blue circles for the \emph{undetectable} MSPs.
Although the match is not perfect, the simulated distribution of \emph{detectable} MSPs
follows the observed distribution better than the distribution based on
\citeauthor{Faucher-Giguere2010}'s approach.
In particular, the problem of generating MSPs with characteristic ages in excess of the age
of the Universe is avoided by construction.

\begin{figure*}[!t]
  \centering
  \includegraphics[width=\textwidth]{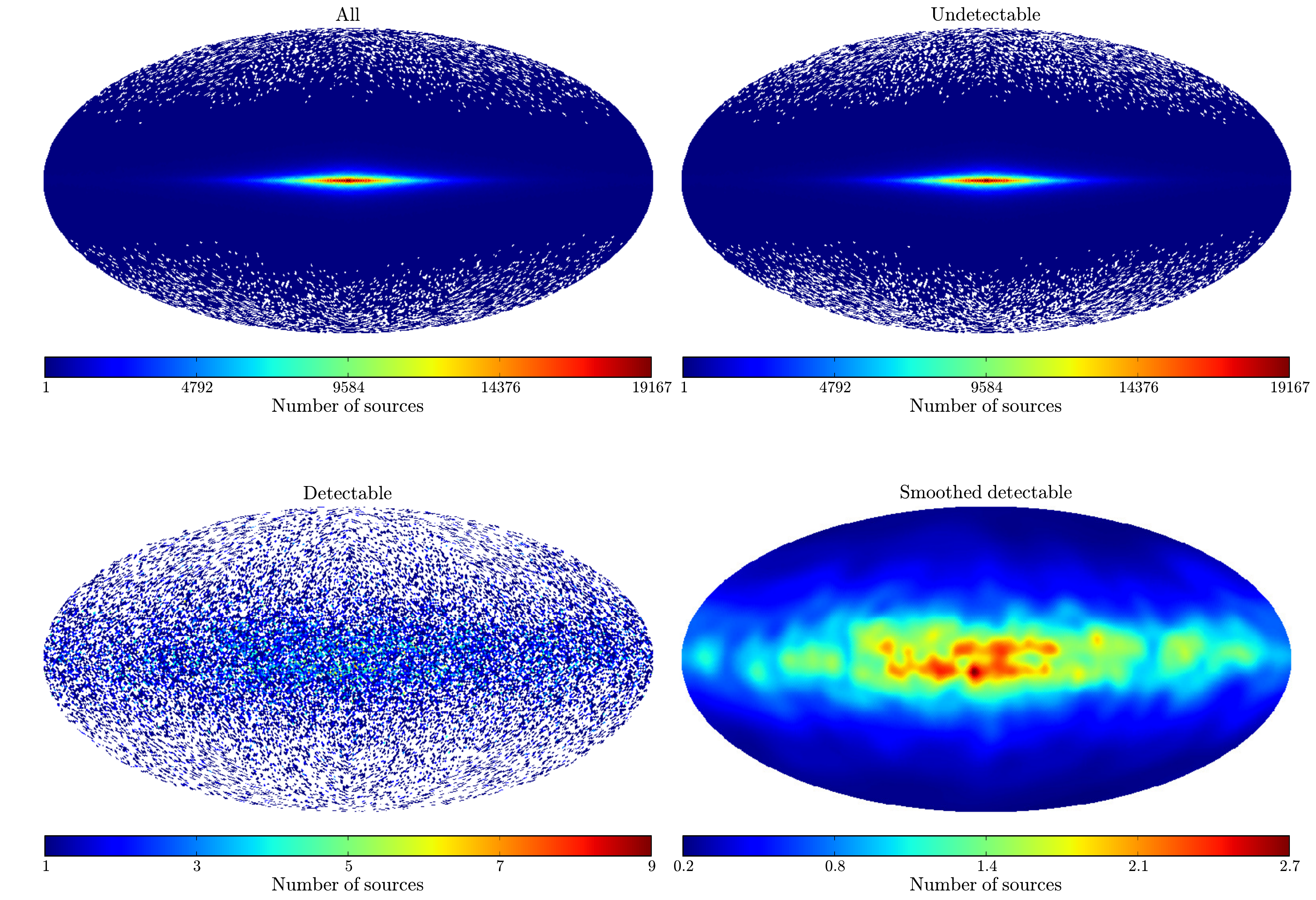}
  \caption{\label{fig:hpm}
    HEALPix maps ($N_{\rm side}=64$) for $R_\mathrm{0}=4.2$ kpc and $\sigma_\mathrm{z} = 500$ pc
    using the \citet{Story2007} luminosity model.
    The panels show
    the entire simulated MSP population ({\em top left}),
    the {\em undetectable} ({\em top right})
    and {\em detectable} ({\em bottom left}) samples,
    and the smoothed {\em detectable} sample ({\em bottom right}).
    }
\end{figure*}

\begin{figure}[!t]
  \centering
  \includegraphics[width=\columnwidth]{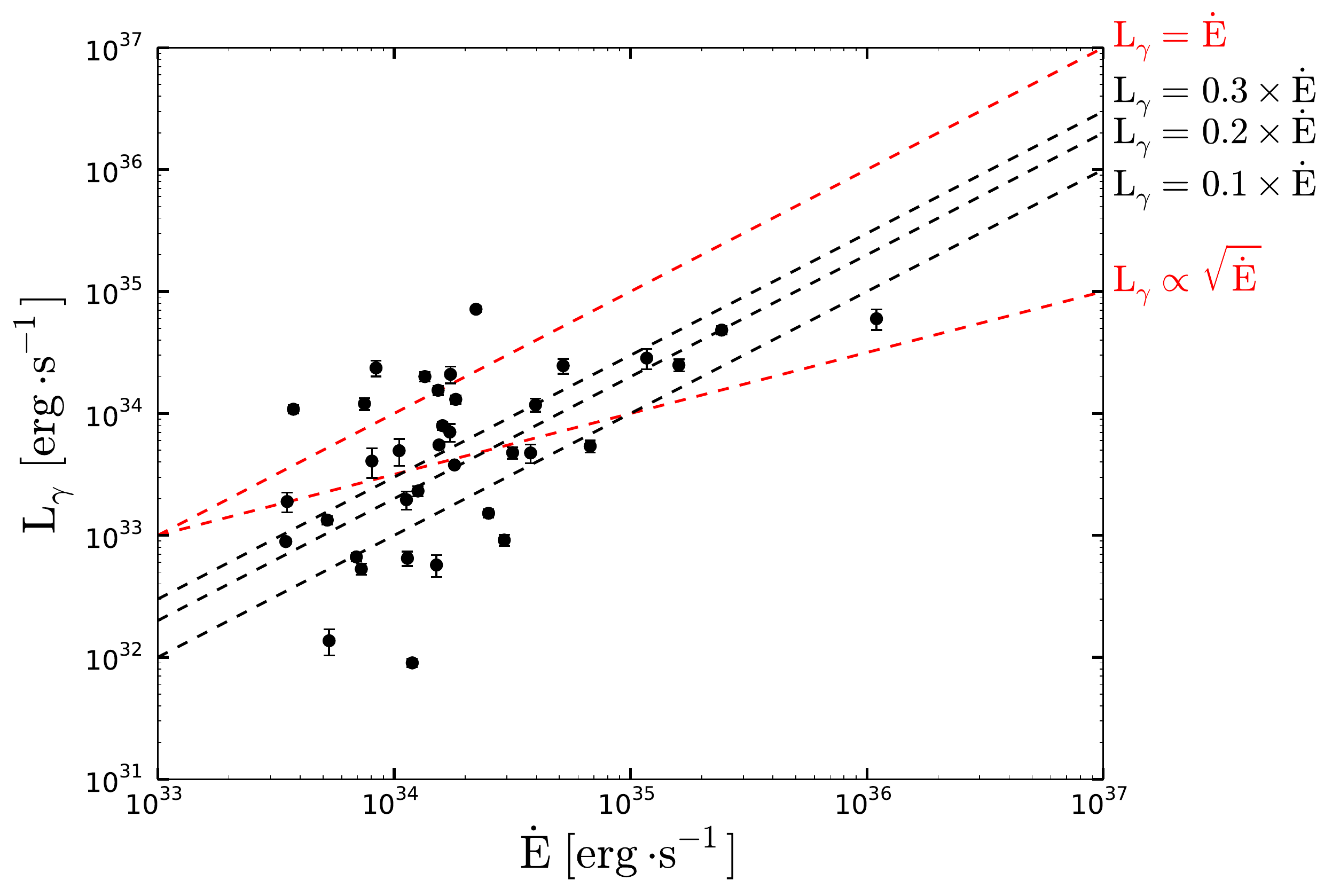}
  \caption{\label{fig:FgFedot}
    Observed \gam-ray luminosity $L_{\gamma}$ versus spin down power $\dot E$ for the
    36 \fermiLAT\ MSPs used in this study.
    The dashed lines indicate empirical laws used to predict the \gam-ray luminosity of individual
    MSPs (see text).
  }
\end{figure}
\subsection{Luminosity function\label{sec:luminosity}}
We use the empirical law
\begin{equation}
  L_{\gamma} = \eta \times \dot E
\label{eq:gammaluminosity}
\end{equation}
to estimate the \gam-ray luminosity from the spin-down luminosity
$\dot E = 4\pi^2 I \dot P / P^3$
of the MSPs,
where $I$ is the moment of inertia (assumed to be equal to 10$^{45}$~$\mathrm{g \, cm^2}$),
$P$ the spin period, $\dot P$ the period derivative of the pulsar
and $\eta$ a proportionality factor.
The use of Eq.~(\ref{eq:gammaluminosity}) is motivated by the fact that the observed
\gam-ray luminosities for the \fermiLAT\ MSPs follow a rough proportionality with
$\dot E$.
This relation is illustrated in Figure~\ref{fig:FgFedot}, which shows the \gam-ray luminosity measured
by \fermiLAT\  for the 36 MSPs used in this work as a function of $\dot E$.
The luminosities $L_{\gamma}$ have been derived from the \gam-ray fluxes
$F_{\gamma}$, which were observed by \fermiLAT\ using Eq.~(\ref{eq:flux}) and 
were consistent with $\Delta \Omega=4 \pi$.
For clarity, the displayed error bars reflect only the uncertainties in the estimation of the 
\gam-ray flux and do not include uncertainties in pulsar distance.
Uncertainties in distance are substantial and dominate the overall error budget, which may
explain a significant fraction of the observed dispersion.
Orientation effects for individual pulsars also play very likely an important role.

We show the luminosity laws (Eq.~\ref{eq:gammaluminosity}) for
$\eta=0.1, 0.2, 0.3,$ and $1.0$ as dashed lines.
We also show the relation $L_{\gamma} \propto \sqrt{\dot E}$
that has been advocated by some authors in the past \citep[e.g.][]{Faucher-Giguere2010}.
Notably, $\eta=0.2$ presents a reasonable average value, and we adopt this
value for our study.
However, we explore the impact of this choice by varying $\eta$ between $0.05$
(which is the value used by \citet{Faucher-Giguere2010}) and $0.3$
(see section \ref{sec:LumLaw}).

\section{Constraining the Galactic MSP population \label{sec:const}}

\subsection{Maximum likelihood method\label{sec:lkhd}}

We now use our MSP population model to derive constraints on the Galactic population of
\gam-ray emitting MSPs.
We use a maximum likelihood ratio method to compare the distribution of MSPs observed
by \fermiLAT\  to our model.
For this purpose, we decompose the sky using a
HEALPix\footnote{\texttt{http://healpix.jpl.nasa.gov}} 
pixelization with $N_{\rm side}=64$
\citep[see][]{Gorski2005}, 
and denote by $n_i$ the number of MSPs observed by the \fermiLAT\ within a given pixel. 
The estimated number of detectable MSPs predicted by our model for each pixel is given by
$e_i$.
The absolute numbers $e_i$ relate to the total number $N_{\rm MC}$ of pulsars that
have been simulated and need to be adjusted by a scaling factor $\alpha$ to match
the data.
The total number of \gam-ray emitting MSPs in our Galaxy is then directly given by
\begin{equation}
N_{\rm exp} = \alpha N_{\rm MC} \, .
\label{eq:nexp}
\end{equation}
The adjustment of $\alpha$ is done by maximizing the log likelihood \citep{Cash1979}
\begin{equation}
  \ln (\mathcal{L}) = \sum \limits_{{i=0}}^{N_{pix}} \left[ n_i \ln(\alpha e_i) - \alpha e_i - \ln(n_i!) \right] 
  \label{eqn:lkhd}
\end{equation}

We also use Eq.~(\ref{eqn:lkhd}) to determine the spatial parameters $\sigma_{\rm r}/R_0$ and 
$z_0$ that best reproduce the distribution of MSPs observed by \fermiLAT.
For this purpose, we compute $\ln (\mathcal{L})$ on a grid spanned by
$\sigma_{\rm r}$ and $z_0$ for the Gaussian density profile (Eq.~\ref{eq:gaussspatial})
or
$R_0$ and $z_0$ for the exponential density profile (Eq.~\ref{eq:expspatial}).
We then use Wilks theorem \citep{Wilks1938}, which states that 
$2 \Delta \ln (\mathcal{L})$ follows a $\chi^2_p$ law with $p$ degrees of freedom,
to derive confidence contours in both parameters.
For our case ($p=2$), a value of $2 \Delta \ln (\mathcal{L})$ equal to
2.30, 6.16, and 11.83
corresponds to 1, 2, and 3 $\sigma$ confidence levels respectively.

Instead of directly using the contours from the grid, we adjust ellipses to the $2 \Delta \ln (\mathcal{L})$
values as a function of radial and vertical scaling parameters by using a moment method
\citep{Teague1980}.
This approach makes our results robust against statistical fluctuations that arise from the limited number
of pulsars drawn in our population model.
We then use these ellipses to derive statistical uncertainties in both scaling parameters.

Figure \ref{fig:hpm} shows HEALPix maps of a population of $N_{\rm MC}=10^7$ MSPs.
All pulsars (top left), the undetectable pulsars (top right), and the detectable
pulsars (bottom left) are presented.
The number of (detectable) MSPs becomes rather sparse with increasing Galactic latitude,
and eventually, some of the pixels in the HEALPix map for the estimate ($e_i$) may become zero,
which makes Eq.~(\ref{eqn:lkhd}) undefined.
We thus applied a modest smoothing to our map of detectable MSPs using an adaptive Gaussian
filter.
For each pixel in our HEALPix map, we adjusted the smoothing kernel of the Gaussian, so that at least
50 MSPs are covered by the kernel.
This drastically reduces the statistical noise in our HEALPix maps, and avoids any empty pixel in our
analysis.
The result of this smoothing operation in illustrated in the bottom right panel of Fig.~\ref{fig:hpm}.

\subsection{Sensitivity map \label{sec:sensitivity}}
\begin{figure}[!t]
  \centering
  \includegraphics[width=\columnwidth]{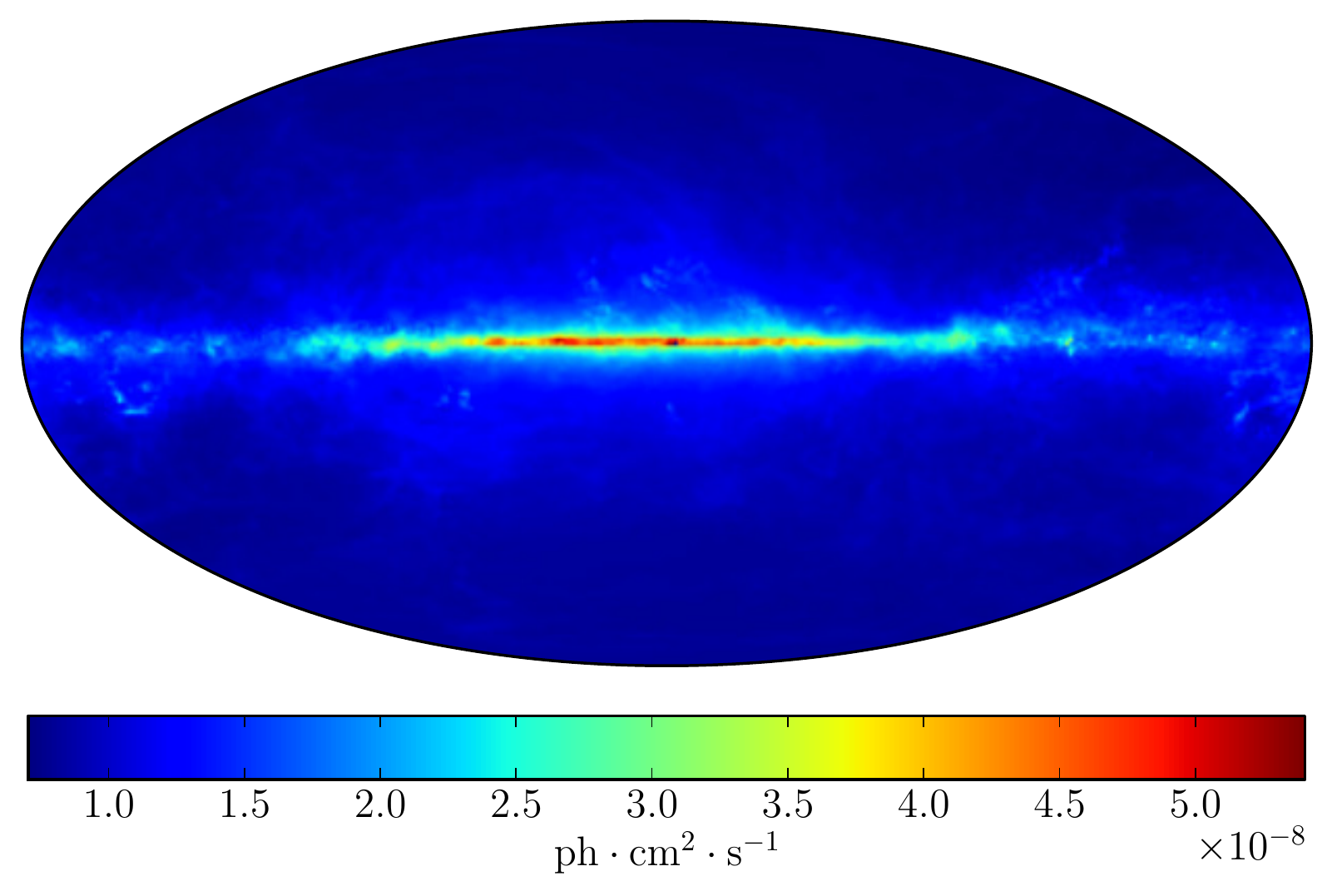}
  \caption{\label{fig:sensitivity1FGL}
  Sensitivity map of the \fermiLAT\ for detection of 1FGL point sources, which are given
  in Galactic coordinates and Aitoff projection \citep{Abdo2010a}.
  }
\end{figure}

An essential ingredient for the separation of the detectable from the undetectable population
of Galactic MSPs is the sensitivity map of the \fermiLAT\ shown in figure~\ref{fig:sensitivity1FGL}.
This sensitivity map is an estimate of the minimum flux that a point source needs to pass to be
detected by the telescope and to be integrated in the \fermiLAT\ source catalogue.
Our sample of MSPs used in this work were all detected during the first year of \fermiLAT\ operations,
and are all included in the 1FGL source catalogue \citep{Abdo2010a}.
We thus use the corresponding sensitivity map that was published with the 1FGL
catalogue.
Instrument and pipeline responses are implicitly included in this estimation, and some assumptions 
about the spectral shape of the sources (power law with index $2.2$), instrumental backgrounds, and astrophysical
backgrounds have been made \citep[see Appendix A of][]{Abdo2010a}.
We discuss the dependency of our results on the sensitivity map in a dedicated section~\ref{sec:uncert}.

\begin{figure*}[!ht]
  \centering
  \subfigure[\stexp\label{sfiga:result}]{\includegraphics[width=0.45\textwidth]{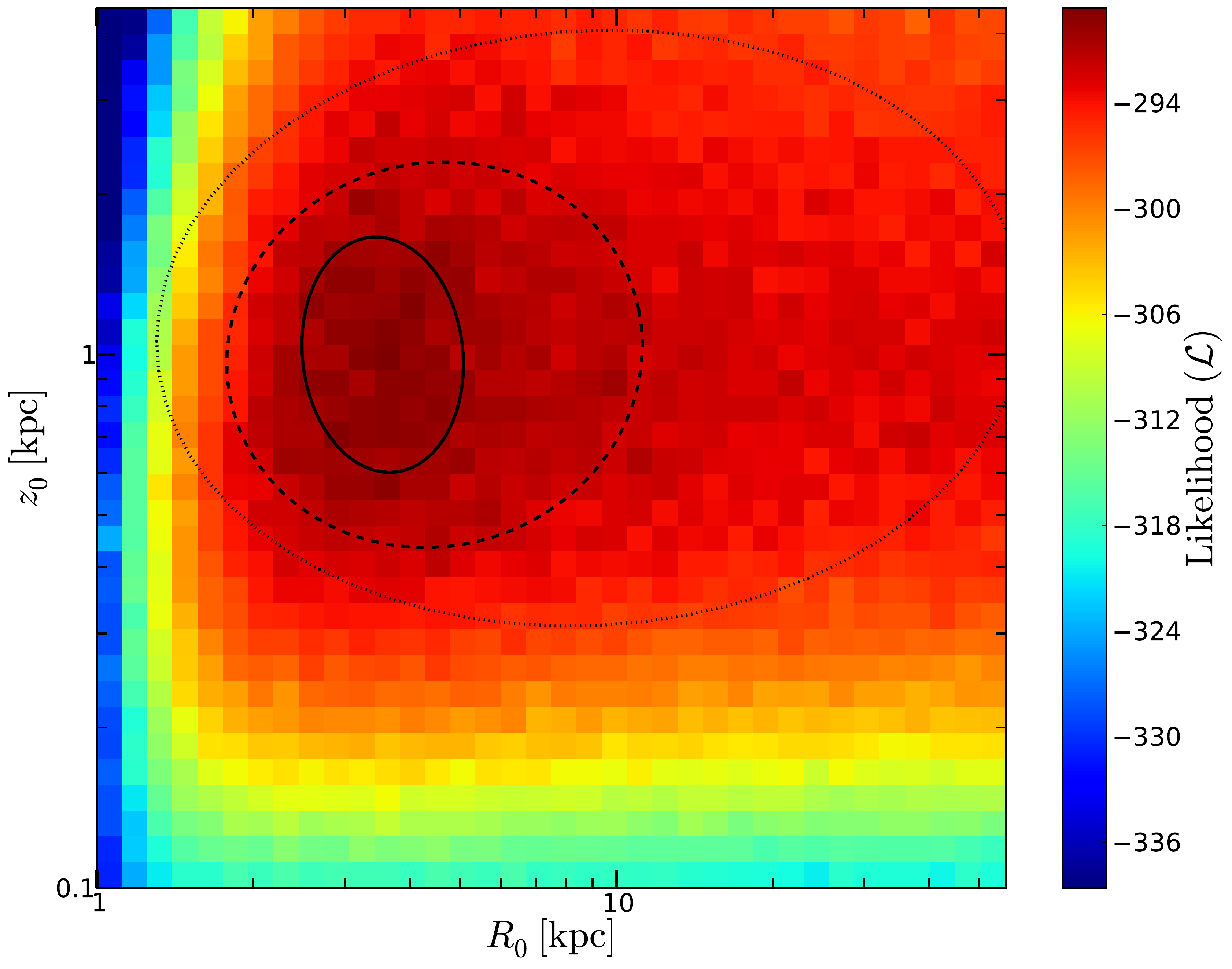}}
  \subfigure[\stgau\label{sfigb:result}]{\includegraphics[width=0.45\textwidth]{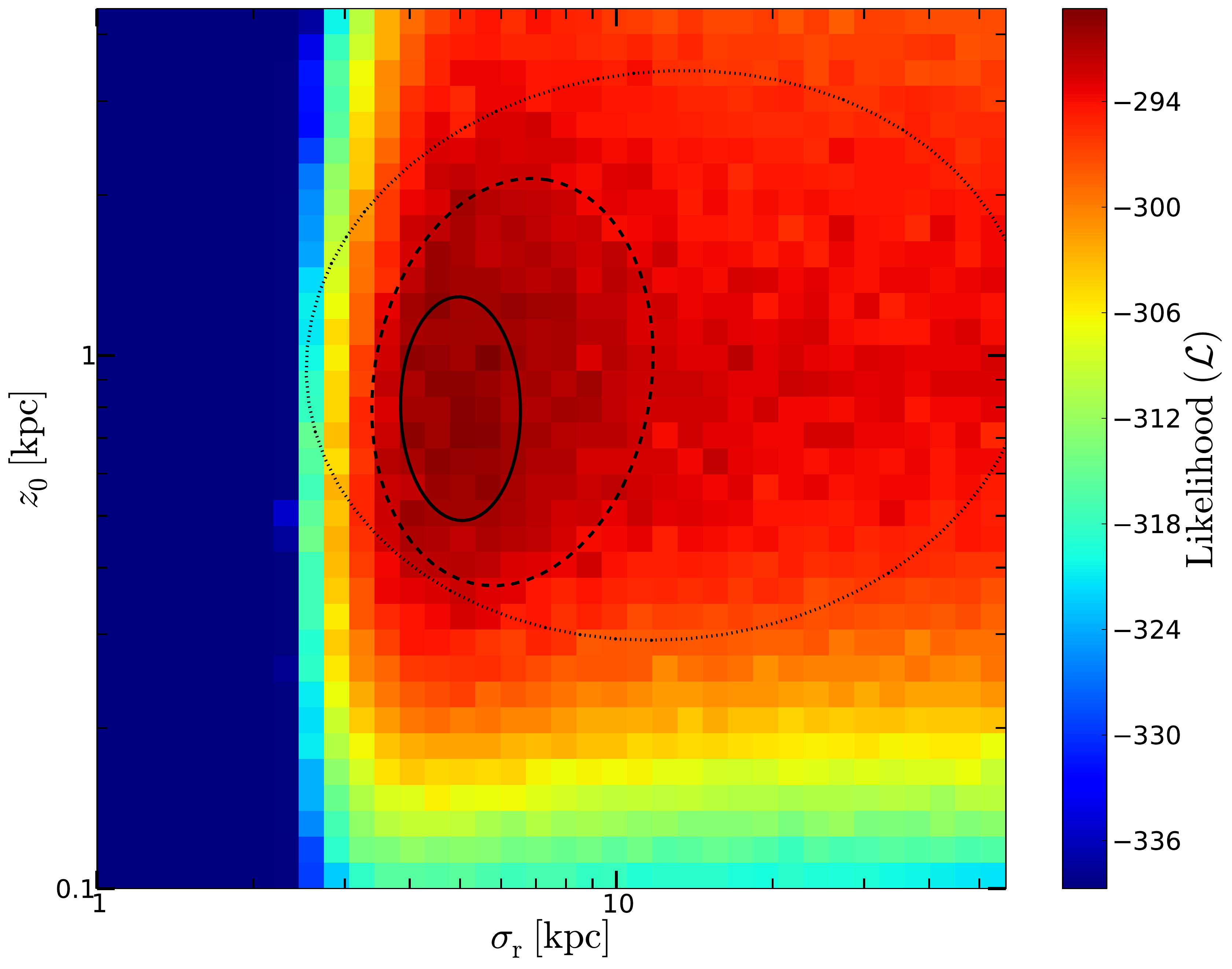}}
  \subfigure[\fgexp\label{sfigc:result}]{\includegraphics[width=0.45\textwidth]{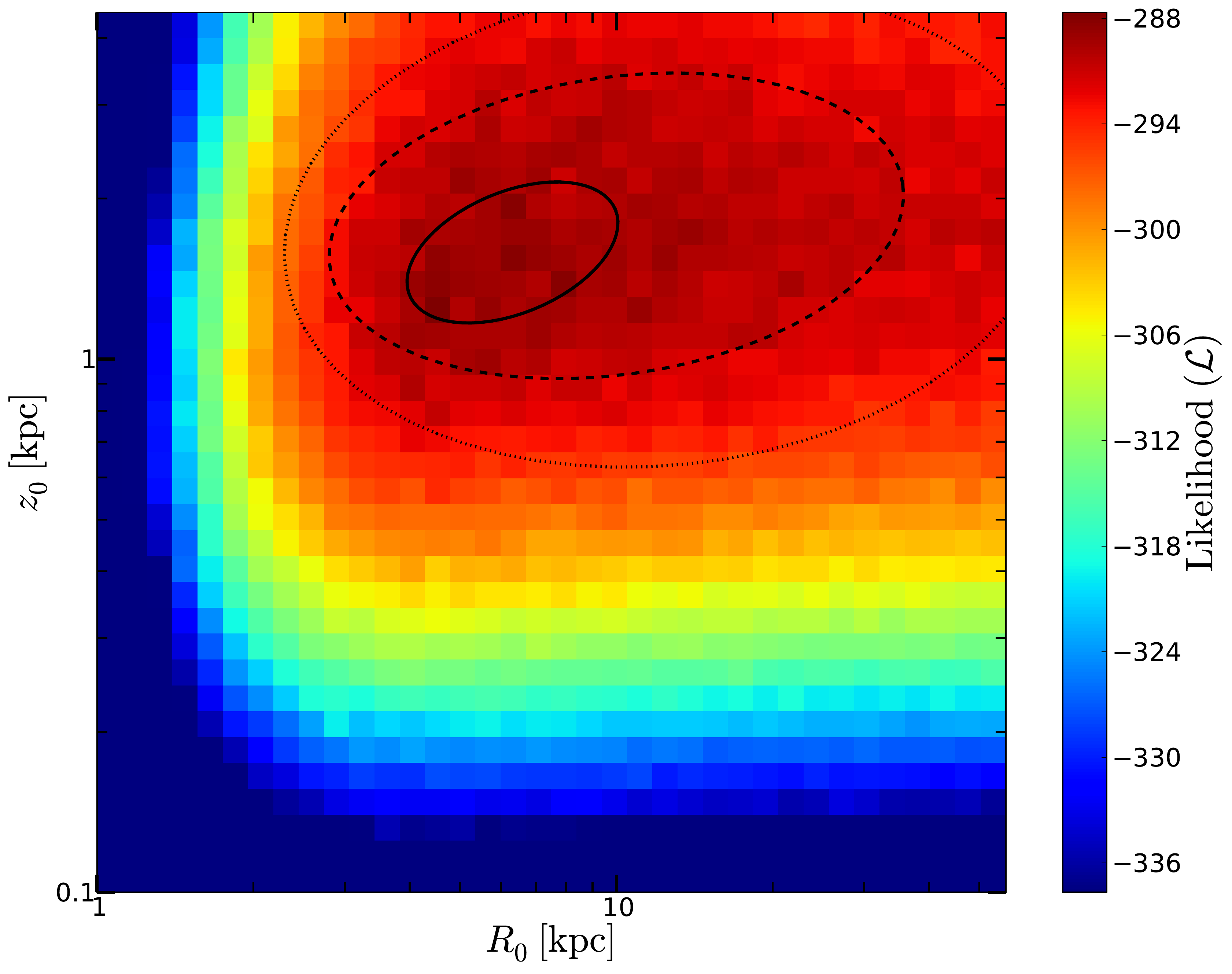}}
  \subfigure[\fggau\label{sfigd:result}]{\includegraphics[width=0.45\textwidth]{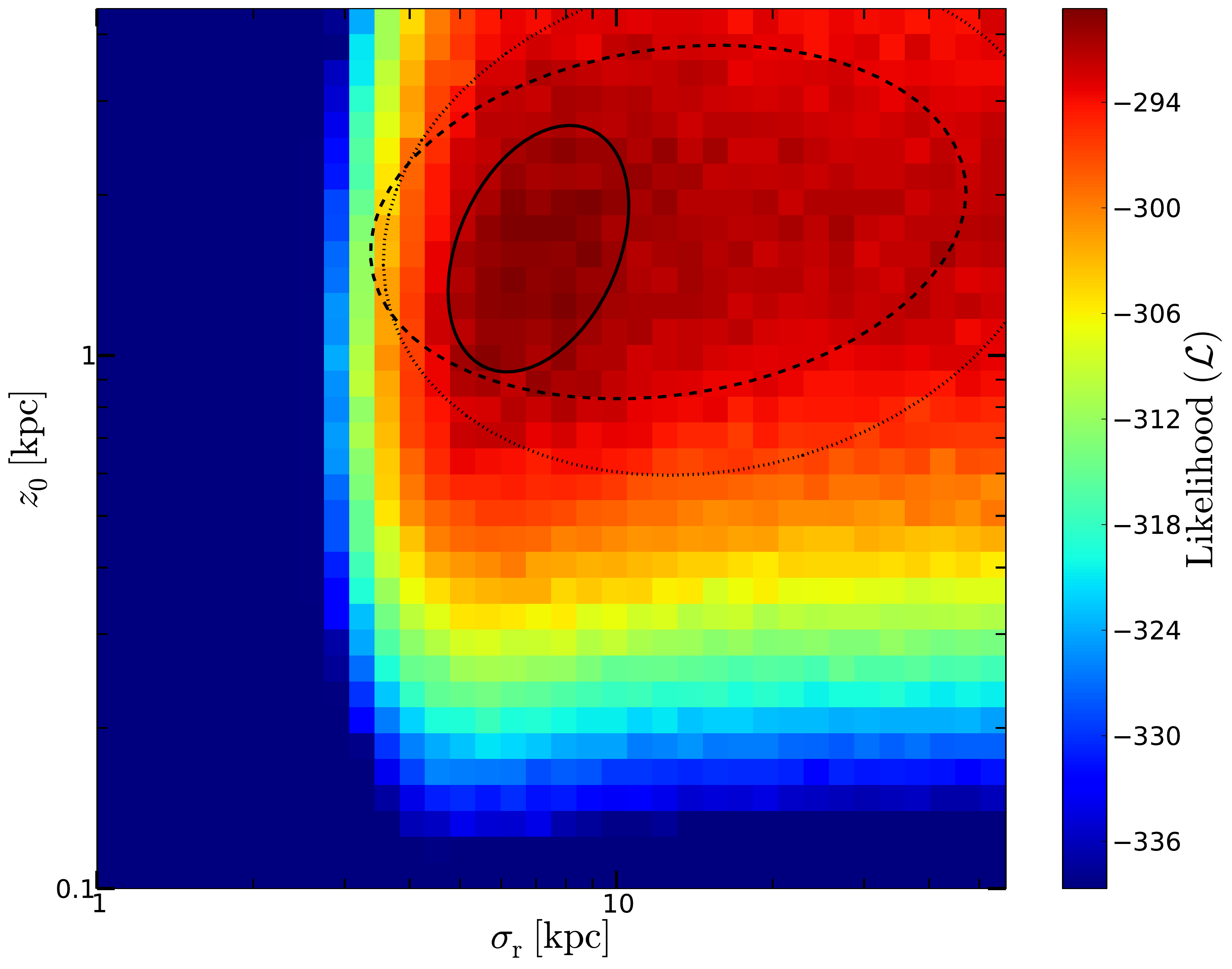}}
  \caption{\label{fig:result}
  Log-likelihood as a function of radial scale length ($R_0$ or $\sigma_r$) and vertical scale
  height ($ z_0$) for four MSP population models (see text).
  Contours show the 1, 2, and $3\sigma$ confidence ellipses that have been computed
  from the log-likelihood maps (colours).
  All maps show the same dynamic range in log-likelihood.
    }
\end{figure*}

\subsection{Results\label{sec:res}}

The results of the maximum likelihood analysis are illustrated in Fig.~\ref{fig:result} and summarized
in Table \ref{tab:result}.
For both $P-\dot P$ modelling approaches
  (designated by ST for \citet{Story2007} and FG for \citet{Faucher-Giguere2010}),
we explored the exponential density profile (Eq.~\ref{eq:expspatial}; models \stexp\ and \fgexp)
and the Gaussian density profile (Eq.~\ref{eq:gaussspatial}; models \stgau\ and \fggau).
For all of those models, a value of $N_{\rm MC} = 10^7$ pulsars have been simulated.

\begin{table}[!t]
  \renewcommand{\arraystretch}{1.5} 
  \newcommand{\sigRr}{\mbox{$R_\mathrm{0}/\sigma_\mathrm{r}$}}
  \caption{Results of the maximum likelihood analysis}
  \centering
  \begin{tabular}{ccccc}
    \hline\hline
    Model name & \stexp & \stgau & \fgexp & \fggau \\
    Spatial distribution & Exp. & Gauss. & Exp. & Gauss. \\
    \hline
    \sigRr\ (kpc) & $4^{+7}_{-3}$ & $6^{+5}_{-3}$ & $10^{+26}_{-7}$ & $13^{+34}_{-9}$ \\
    $z_\mathrm{0}$ (kpc) & $1.0^{+1.3}_{-0.6}$ & $0.9^{+1.3}_{-0.5}$ & $1.8^{+1.7}_{-0.9}$ & $1.8^{+2.0}_{-0.9}$\\
    \Nexp\ ($10^{3}$) & $11^{+4}_{-4}$ & $9^{+3}_{-3}$ & $4^{+1}_{-1}$ & $4^{+1}_{-1}$ \\
    \Fcr\ ($10^{-7}$)\tablefootmark{*} & $21^{+7}_{-7}$ & $16^{+5}_{-5}$ & $8^{+3}_{-3}$ & $7^{+2}_{-2}$\\
    \Ipsr\ ($10^{-7}$)\tablefootmark{*} & $0.24^{+0.08}_{-0.08}$ & $0.22^{+0.07}_{-0.07}$ & $0.15^{+0.05}_{-0.05}$ & $0.15^{+0.05}_{-0.05}$\\
    \hline
  \end{tabular}
  \tablefoot{\label{tab:result}
  The scaling parameters $R_\mathrm{0}$, $\sigma_\mathrm{0}$, and $z_\mathrm{0}$ are given in units
   of kpc.
   $N_\mathrm{exp}$ is the total number of Galactic MSPs in the Milky Way given in units of $10^3$. 
    \Fcr\ are the estimated integrated average intensity above 100 MeV
    of the undetectable MSPs in units of $10^{-7}$ \iunit\
    from the central Galactic radian, defined by $|l| \leq$ 30\degree\ and $|b| \leq$ 10\degree.
    \Ipsr\ gives the integrated average intensity above 100 MeV
    of the undetectable MSPs in the same units at
    high Galactic latitudes ($|b| \ge 40\degree$).
    All uncertainties are only statistical (2$\sigma$ confidence level). \\
    \tablefoottext{*}{Units are \iunit}
  }
\end{table}

Figure \ref{fig:result} shows the log-likelihood function $\ln (\mathcal{L})$ as a function of the radial
($R_0$ or $\sigma_r$) and vertical ($z_0$) scale parameters for all models.
All panels of the figure have been adjusted to the same dynamic range.
Solid, dashed, and dotted ellipses show 1, 2, and 3$\sigma$ confidence levels, respectively, and the 
colours depict the log-likelihood values.
From the minimum and maximum values reached by the $2\sigma$ confidence
ellipses and the centroid, mean values and statistical uncertainties are derived for the scale parameters
and are summarized in Table \ref{tab:result}.
From the value of the scaling factor $\alpha$ at the ellipse centroid, the expected number
\Nexp\ of gamma-ray emitting MSPs in our Galaxy is derived using Eq.~(\ref{eq:nexp}).
Uncertainties of \Nexp\ are determined by finding the values of $\alpha_{\rm min}$ and
$\alpha_{\rm max}$ for which the log-likelihood function (Eq.~\ref{eqn:lkhd}) decreases by 
a value of 2 with respect to the maximum (corresponding to a $2\sigma$ confidence
interval).

By comparing models \stexp\ to \fgexp\ and models \stgau\ to \fggau, it becomes obvious that
the choice of the $P-\dot P$ model has
some impact on the results, which leads to higher values for the scaling parameters for
{\tt FG} with respect to {\tt ST}.
The detectable pulsars predicted by the {\tt FG} model have an average luminosity that
is larger than those predicted by the {\tt ST} model, which is readily seen in their
$P-\dot P$ distribution shown in Fig.~\ref{fig:PPdot}.
Detectable {\tt FG} pulsars are thus seen at greater distances.
For a given set of scale 
parameters their spatial distributions more concentrated towards the Galactic plane and the 
inner Galaxy with respect to the {\tt ST} model.
Consequently, larger scale parameters are required for the {\tt FG} pulsars to fit to the
observed spatial distribution of MSPs.

In summary, the {\tt ST} models suggest a vertical scale height $z_\mathrm{0}$ of $\sim1$ kpc,
while the {\tt FG} models predict a larger scale height of $\sim1.8$ kpc.
For the radial exponential scale length $R_0$, the \stexp\ model suggests a value of
$\sim4$ kpc, while the \fgexp\ model predicts $\sim10$ kpc.
Using a Gaussian radial profile leads to $\sigma_{\rm r}$ values of $\sim6$ kpc
for \stgau\ and $\sim13$ kpc for \fggau.

In addition to the spatial parameters, our analysis also predicts the total number \Nexp\ of MSPs in
our Galaxy that are potentially detectable in \gam-rays.
Because the {\tt FG} pulsars are on average brighter, a higher fraction of the Galaxy
becomes detectable with that model.
Because the normalization of the model is given by the constraint that the number of pulsars
in the detectable volume is equal to 36, a larger volume implies a smaller scaling factor
$\alpha$.
Consequently, \Nexp\ is lower for the {\tt FG} model ($\sim4\,000$) than for the {\tt ST} 
model ($\sim10\,000$).

Our model also allows the prediction of the total \gam-ray intensity received at Earth from MSPs
that do not pass the detection threshold, hence forming a diffuse background from
unresolved sources \citep{Bhattacharya1991}.
We estimate this intensity $S_\Omega$ by summing the individual fluxes above 100 MeV
of all undetectable MSPs in a given region $\Omega$ divided by the solid area of the region, i.e.
\begin{equation}
  S_\Omega = \frac{1}{\Omega} \sum \limits_{\tiny{\begin{array}{c} (l,b) \in\ \Omega\ \\ F_\gamma < F_{\rm sensitivity} \end{array}}} F_\gamma
\end{equation}
In Table \ref{tab:result}, the expected average intensities above 100 MeV
from the central Galactic radian
(\Fcr, integrated over $|l| \leq$ 30\degree\ and $|b| \leq$ 10\degree)
and high Galactic latitudes
(\Ipsr, integrated over $|b| \ge 40\degree$) are shown.

The diffuse fluxes follow the same trend as the expected number of MSPs in the
Galaxy, with the {\tt ST} model predicting larger fluxes than the {\tt FG} models.
Using the {\tt ST} models, the expected diffuse flux from the Galactic central radial
is estimated to $\sim2 \times 10^{-6}$ \iunit, while this flux is
only about $\sim0.8 \times 10^{-6}$ \iunit for the {\tt FG} models.
This result corresponds to a total of $1\%$ or less of the Galactic \gam-ray flux that is observed 
by \fermiLAT\ from that area \citep{Ackermann2012a}.
At high Galactic latitudes, the {\tt ST} models predict a value of 
$\sim2.0 \times 10^{-8}$ \iunit for an average \gam-ray intensity 
above 100 MeV due to unresolved MSPs, while
the {\tt FG} models predict a value of $\sim1.5 \times 10^{-8}$ \iunit.
This intensity corresponds to an approximate value of a few permil of the extragalactic background 
intensity \citep{Abdo2010EB}.

%

\begin{figure*}[!ht]
  \centering
  \subfigure[\stmsp\label{sfiga:una}]{\includegraphics[width=0.45\textwidth]{figures/plot-lkhd_sim_story_opt.pdf}}
  \subfigure[\stuna\label{sfigb:una}]{\includegraphics[width=0.45\textwidth]{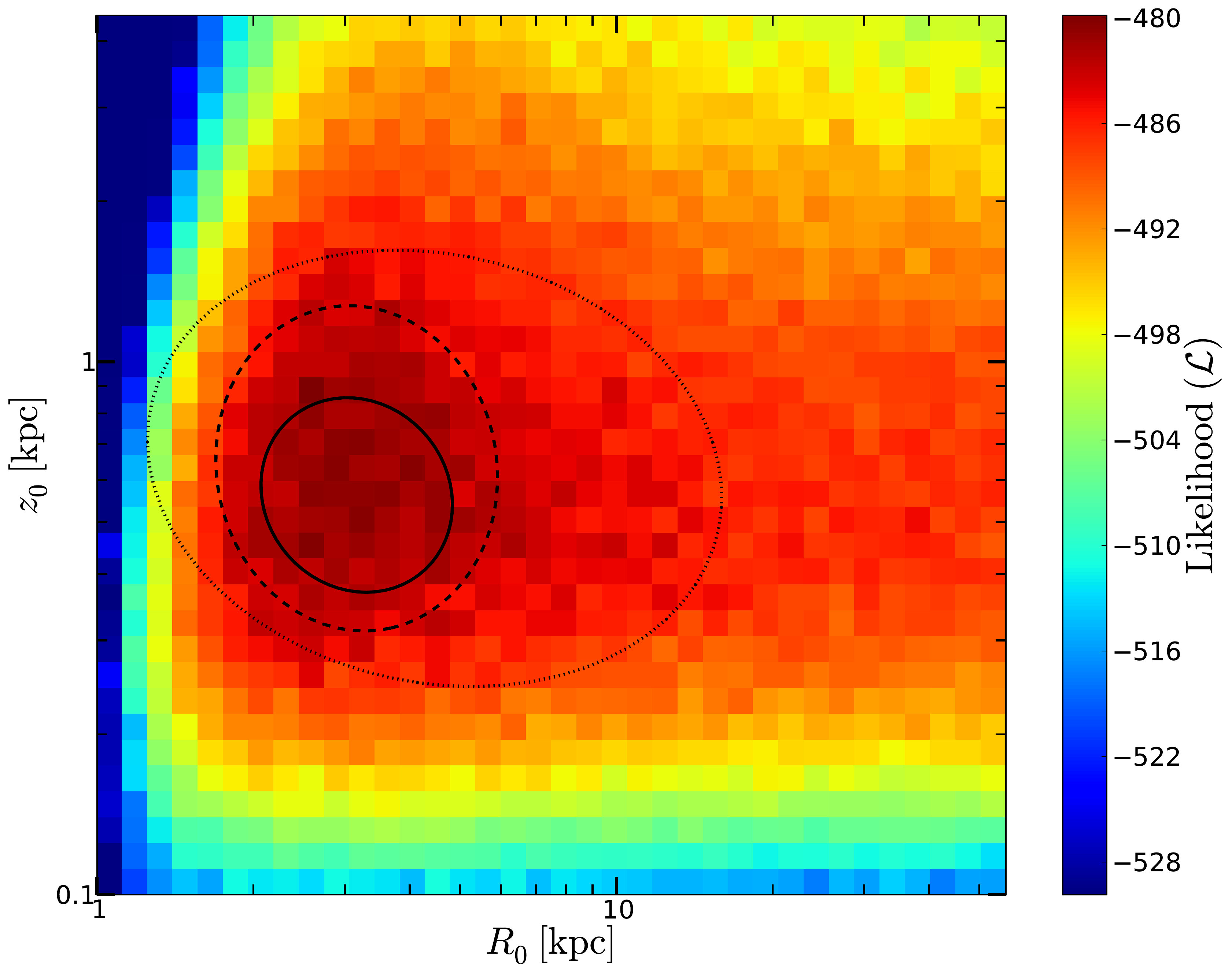}}
  \subfigure[\stst\label{sfigc:una}]{\includegraphics[width=0.45\textwidth]{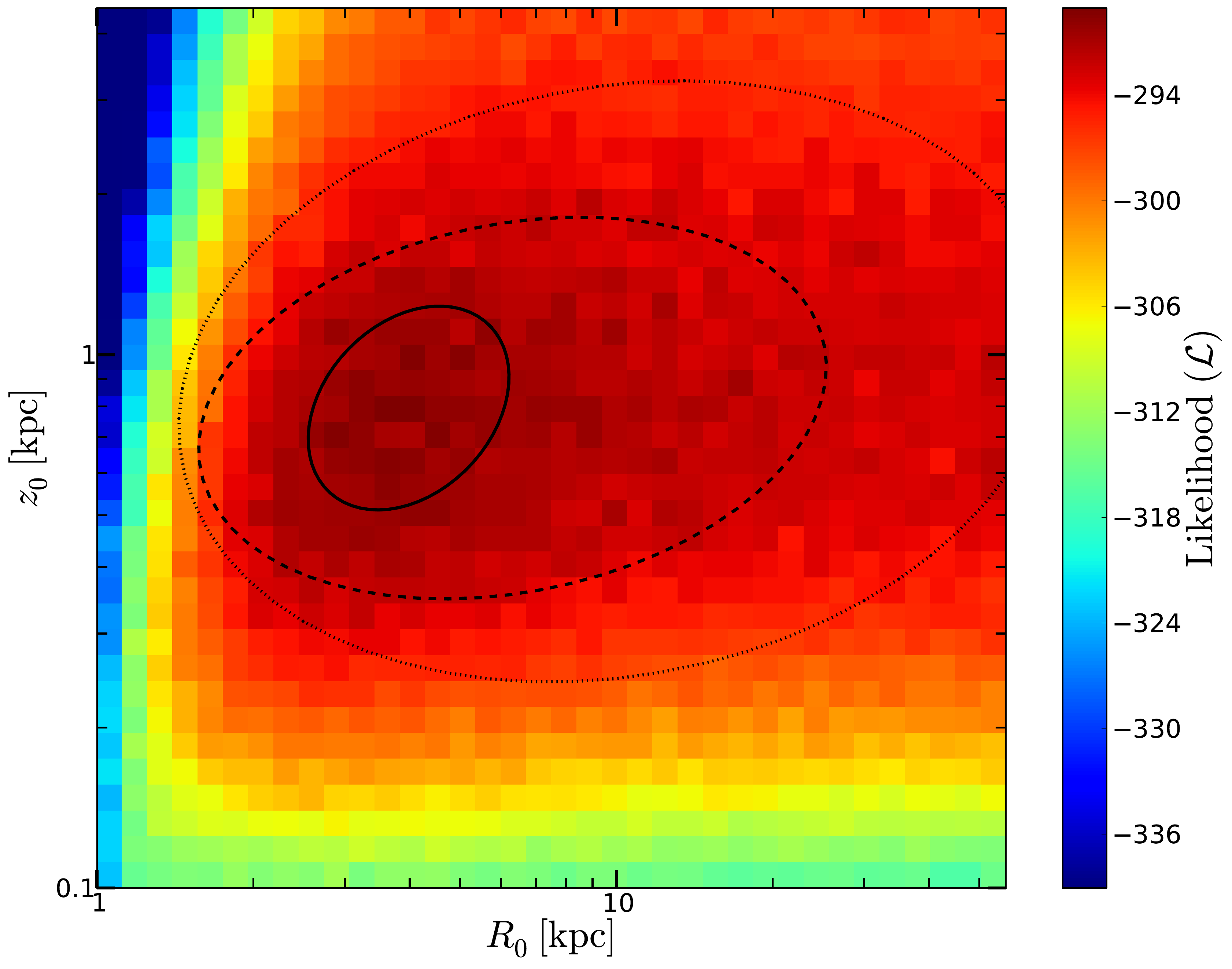}}
  \subfigure[\stsh\label{sfigd:una}]{\includegraphics[width=0.45\textwidth]{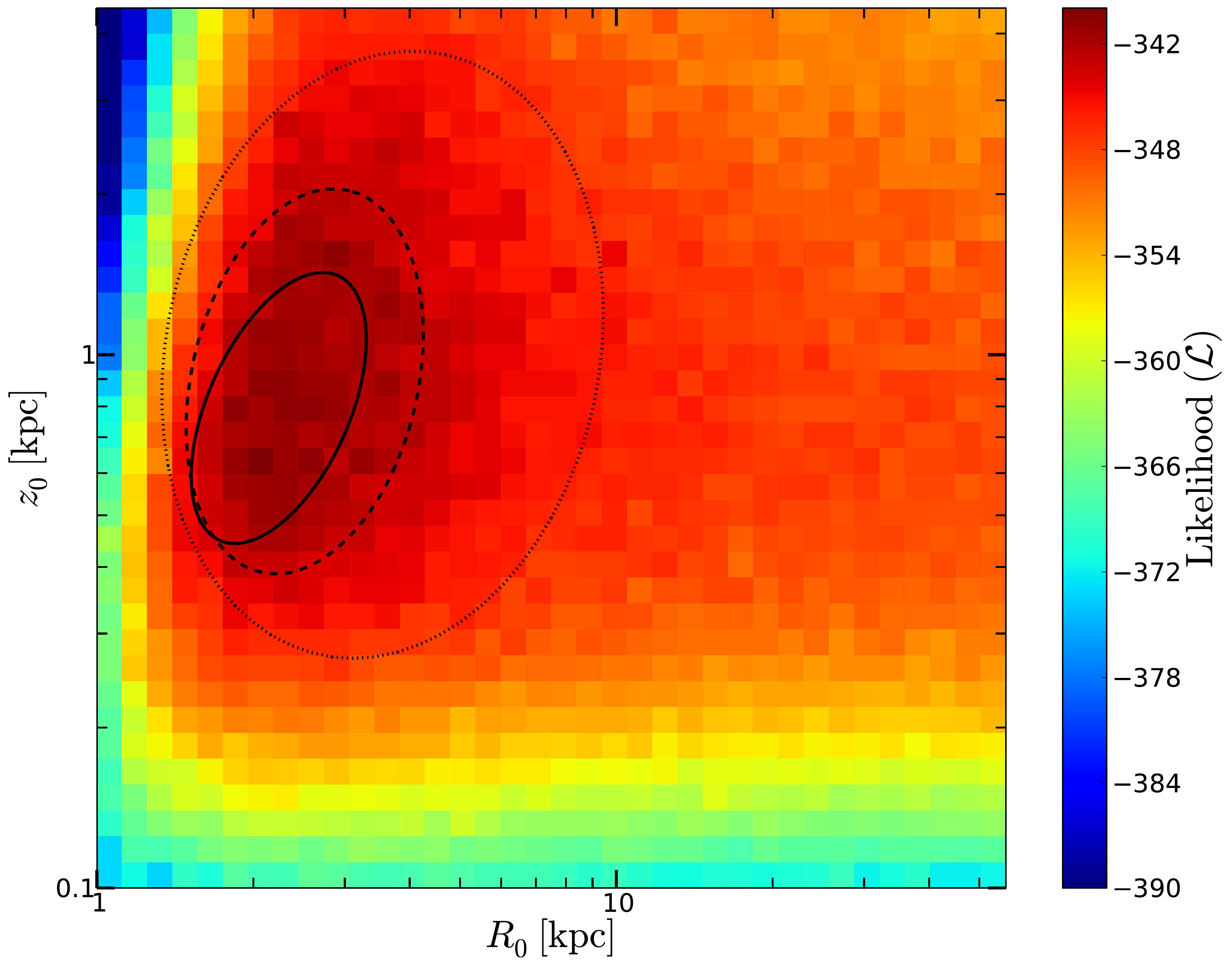}}
  \caption{\label{fig:una}
  Log-likelihood as a function of radial scale length ($R_0$ or $\sigma_r$) and vertical scale
  height ($ z_0$) for the model 
  \stexp.
  The panels show (from top left to bottom right) the results for 
  \Nmsp\ MSPs (a),
  \Nuna\ MSP candidates (b),
  \Nmsp\ mock MSPs (c), and
  \Nuna\ mock MSPs (d).
  Contours show the 1, 2, and $3\sigma$ confidence ellipses that have been computed
  from the log-likelihood maps (colours).
  All maps show the same dynamic range in log-likelihood.
  }
\end{figure*}

\subsection{Extended MSP sample\label{sec:prospect}}
So far, we applied our analysis to the sample of \Nmsp\ MSPs that have been detected in the
1FGL catalogue and for which \gam-ray pulsations have been significantly detected.
Through searches with radio telescopes for pulsations from the error regions of yet unassociated
\fermiLAT\ sources that obey spectral and temporal characteristics that are reminiscent of MSPs,
many of the MSPs in our sample have only been identified as such after the publication of the
catalogue.
Further unassociated sources in 1FGL show similar characteristics.
Based on the classification analysis of unassociated sources made by \citet{Ackermann2012a},
we extend our sample by including unassociated 1FGL sources that were classified in this work
as potential pulsars and exclude all sources that are either flagged as dubious (flag c) or that are
situated closer than $|b| \le 10\degree$ to the Galactic plane.
This list provides 38 additional sources that eventually could be \gam-ray emitting MSPs.
Adding to this list of the known MSPs, where we also excluded all objects
situated closer than $|b| \le 10\degree$ to the Galactic plane for consistency, we obtain a final list of
\Nuna\ potential \gam-ray emitting MSPs, as seen in the 1FGL catalogue.

Figure~\ref{sfigb:una} shows the resulting log-likelihood distribution when
using this sample instead of the \Nmsp\ MSPs that were used.
We excluded in this analysis all HEALPix pixels with pixel centres of $|b| \le 10\degree$.
We only show here results for model \stexp.
For reference, we also show the log-likelihood distribution obtained for the \Nmsp\ 
MSPs for model \stexp\ in Fig.~\ref{sfiga:una}.

With the additional MSP candidates, the Galactic scaling parameters are now much 
better constrained.
Our model predicts 
a radial scale length of $R_0 \sim 3$ kpc,
a vertical scale height of $z_0 \sim 0.6$ kpc,
and a total number of $\Nexp \sim 22\,000$ \gam-ray emitting MSPs in the Milky Way.
This number is about a factor of $\sim2$ larger than our estimate that is based on the 36 confirmed
MSPs, which directly reflects the larger number of MSPs in the extended sample.
As the extended sample includes all objects of the 1FGL catalogue, which shows the spectral and
temporal characteristics of MSPs, it may include some sources that simply mimic MSPs, and
in this sense, the results obtained for the extended sample may be interpreted as upper
limits.
However, some unassociated sources in 1FGL may be too faint to exhibit a significant
spectral curvature, which is required in the extended sample.
The true number of MSPs in the 1FGL catalogue may thus still be a bit larger.

Using the extended sample, the
unresolved average background intensity above 100 MeV from the Galactic central radian 
is now estimated to be $\Fcr \sim 6 \times 10^{-6}$ \iunit\ (about 2\% of the observed Galactic
diffuse flux in that region), and
the average intensity above 100 MeV at high Galactic latitudes is predicted to be
$\Ipsr \sim 2 \times 10^{-8}$ \iunit\ (about 0.4\% of the observed intensity).

\subsection{Possible biases\label{sec:uncert}}

\subsubsection{Statistical variations}
Ideally, our results should be driven by the particular sample of observed MSPs (or MSP candidates)
that we have at hand, and the confidence contours should reflect the statistical uncertainties that
arise from the limited number of objects.
We now use mock catalogues of observed MSPs to investigate how statistical variations impact
our results, and to search for possible biases in our analysis.

As a first test, we produce mock MSP catalogues using the bootstrap method \citep{Efron1979}.
In this method, we create mock catalogues by randomly selecting \Nmsp\ objects from the \Nmsp\ 
observed MSPs, which allows individual MSPs to be selected multiple times.
This result leads to a resampling of the list of observed MSP that should be a statistical variation of the
original sample.
We created ten bootstrap samples, and for each sample, we perform a maximum likelihood analysis 
from which we determine the confidence ellipses in the scaling parameters.
We find that the scaling parameters vary substantially between each individual bootstrap samples
and remain compatible within the statistical uncertainties with the values found for the
observed MSP sample.
We thus conclude that the particular shape of the log-likelihood contours is primarily driven
by the statistics of the sample.

Alternatively to the bootstrap analysis, we also performed the maximum likelihood analysis on
subsets of only 30 MSPs that were drawn randomly from the original list of \Nmsp\ objects.
The results of the subset analysis were qualitatively comparable to the bootstrap analysis,
which confirm that the confidence contours are driven by the statistical variations.

\begin{table}
  \renewcommand{\arraystretch}{1.5} 
  \caption{Results with increased sample and simulated samples}
  \centering
  \begin{tabular}{ccccc}
    \hline\hline
    Model name & \stmsp & \stst  & \stuna & \stsh \\
    MSPs & \Nmsp & \Nmsp & \Nuna & \Nuna \\
    Data & Obs. & Sim. & Obs. & Sim. \\
    \hline
    $R_\mathrm{0}$ (kpc) & $4^{+7}_{-3}$ & $6^{+19}_{-5}$ & $3^{+3}_{-1}$ & $3^{+2}_{-1}$ \\
    $z_\mathrm{0}$ (kpc) & $1.0^{+1.3}_{-0.6}$ & $0.8^{+1.0}_{-0.4}$ & $0.6^{+0.6}_{-0.3}$ & $0.9^{+1.2}_{-0.5}$ \\
    \Nexp\ ($10^{3}$) & $11^{+4}_{-4}$ & $10^{+4}_{-4}$ & $22^{+14}_{-14}$ & $29^{+8}_{-8}$ \\
    \Fcr\ ($10^{-7}$)\tablefootmark{*} & $21^{+7}_{-7}$ & $17^{+7}_{-7}$ & $63^{+40}_{-40}$ & $81^{+23}_{-23}$ \\
    \Ipsr\ ($10^{-7}$)\tablefootmark{*} & $0.24^{+0.08}_{-0.08}$ & $0.18^{+0.07}_{-0.07}$ & $0.2^{+0.1}_{-0.1}$ & $0.4^{+0.1}_{-0.1}$ \\
    \hline
  \end{tabular}
  \tablefoot{\label{tab:una}
  Notations are similar to Table \ref{tab:result}.
  The row labelled {\em Data} indicates whether the results are obtained using the observed
  sample of MSPs (Obs.) or whether they are based on a simulation (Sim.).
  }
\end{table}

As a third test, we also performed the maximum likelihood analysis on mock samples composed
of \Nmsp\ MSPs which were generated randomly from our population model.
In total, we performed 20 mock analyses. We show one result in Fig.~\ref{sfigc:una}
for illustration.
The parameter values obtained for this simulation are given in the third column of Table~\ref{tab:una}.
The results for this specific mock sample are reasonably close to the results obtained 
using the observed \Nmsp\ MSPs (cf.~Fig.~\ref{sfiga:una}), which illustrate that at least one out of 20
samples (i.e.~5\%) produces confidence contours that are compatible with those of the observed
MSPs.
We also note that the 2$\sigma$ confidence contours enclose the true values of 
$R_0=4.2$ kpc and $z_0=500$ pc that were used in the simulation, which confirms
that our parameter estimates are reliable.
This finding is important, because it demonstrates that our analysis procedure does not introduce any
bias in the scaling parameters of the Galactic MSP density distribution.
In particular, the observed MSP sample seems to indicate a larger vertical scale height than
the $500$~pc that have been chosen for the simulation, suggesting that the Galactic latitude
distribution of \gam-ray emitting MSPs may eventually be broader than previously assumed
\citep{Story2007}.

Finally, we also performed 20 mock simulations for a population of \Nuna\ observed MSPs. 
We show one result in Fig.~\ref{sfigd:una}.
The corresponding parameter values are quoted in the fifth column of Table~\ref{tab:una}.
These simulations collerate the results found earlier for \Nmsp\ MSPs, and confirm that the
uncertainties in the analysis are compatible because of only statistical uncertainties.

\subsubsection{Impact of sensitivity map}
As mentioned earlier (cf.~section~\ref{sec:sensitivity}), the sensitivity map is an important ingredient
in our analysis that is used to discern detectable from undetectable MSPs in our simulations and that 
has an impact on the parameters of the Galactic MSP population.
For our study, we use the sensitivity map published for the 1FGL catalogue, which is based
on the hypothesis that all sources have a power law spectral shape with an spectral index
of $\Gamma=2.2$.

\begin{figure}[!t]
  \centering
  \includegraphics[width=\columnwidth]{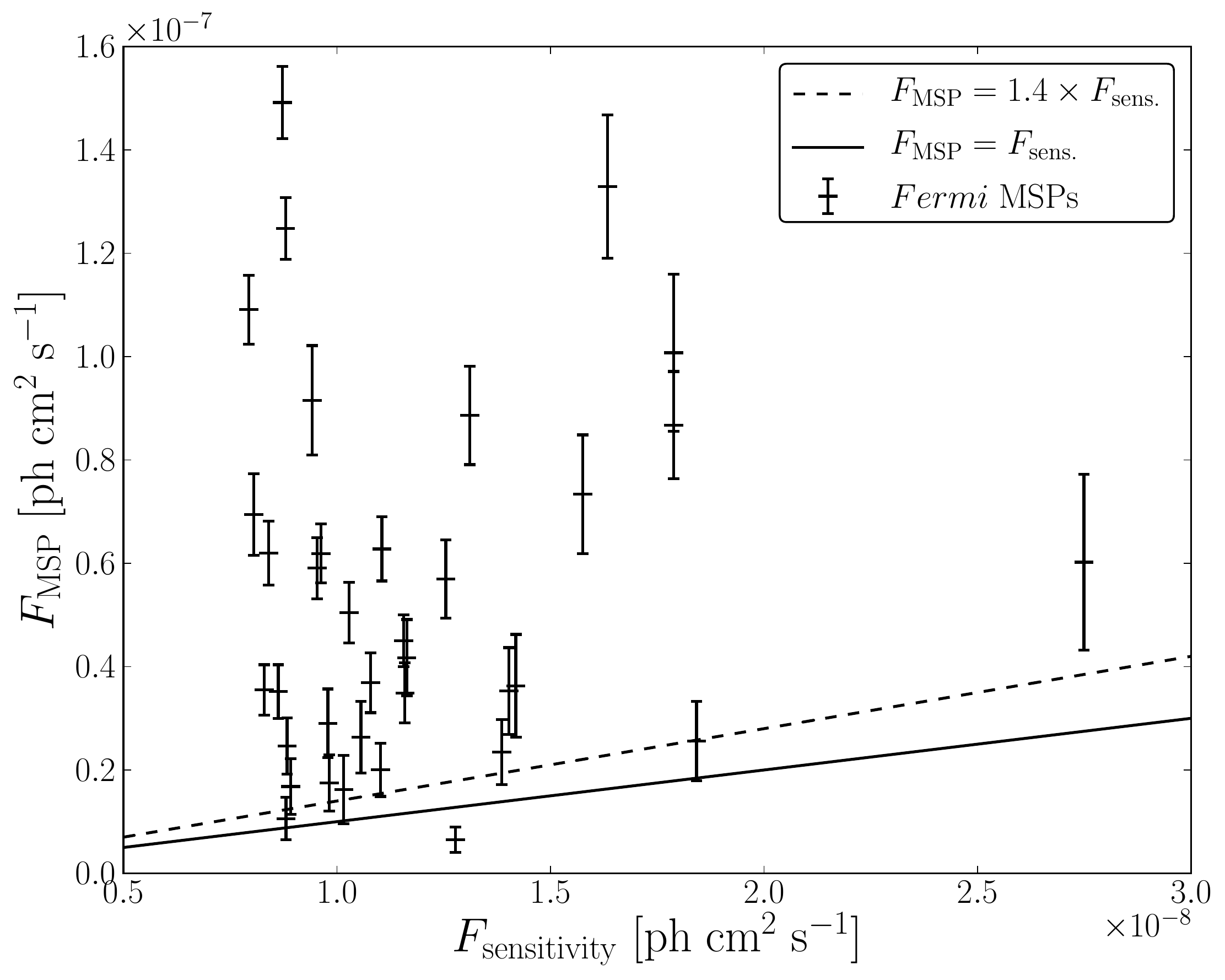}
  \caption{\fermiLAT\ MSPs flux above 100 MeV ($F_\mathrm{MSP}$) 
    against the sensitivity ($F_\mathrm{sensitivity}$). 
    Solid and dashed lines show $F_\mathrm{MSP} = F_\mathrm{sensitivity}$ 
    and $F_\mathrm{MSP} = \sensFact\ \times F_\mathrm{sensitivity}$ respectively.
    }
  \label{fig:plotSensitivity}
\end{figure}

Figure \ref{fig:plotSensitivity} shows the flux above 100 MeV of the \Nmsp\ MSPs used for 
analysis versus the sensitivity value at the respective locations.
Ideally, all MSPs should fall above the solid line, which indicates an MSP flux identical to the
sensitivity value.
We find one object (1FGL J1600.7$-$3055) that violates this constraint, but
detailed inspection of this object reveals an unusually hard spectral power law index 
($\Gamma=1.8$) for this source.
Due to the decreasing size of the point spread function with increasing energy, hard
spectrum sources are in fact easier to detect by \fermiLAT.
The corresponding sensitivity should thus be smaller for such hard spectrum sources
brings the measured flux in agreement with the expected instrument performances.

\begin{figure*}[t!]
    \includegraphics[width= \textwidth]{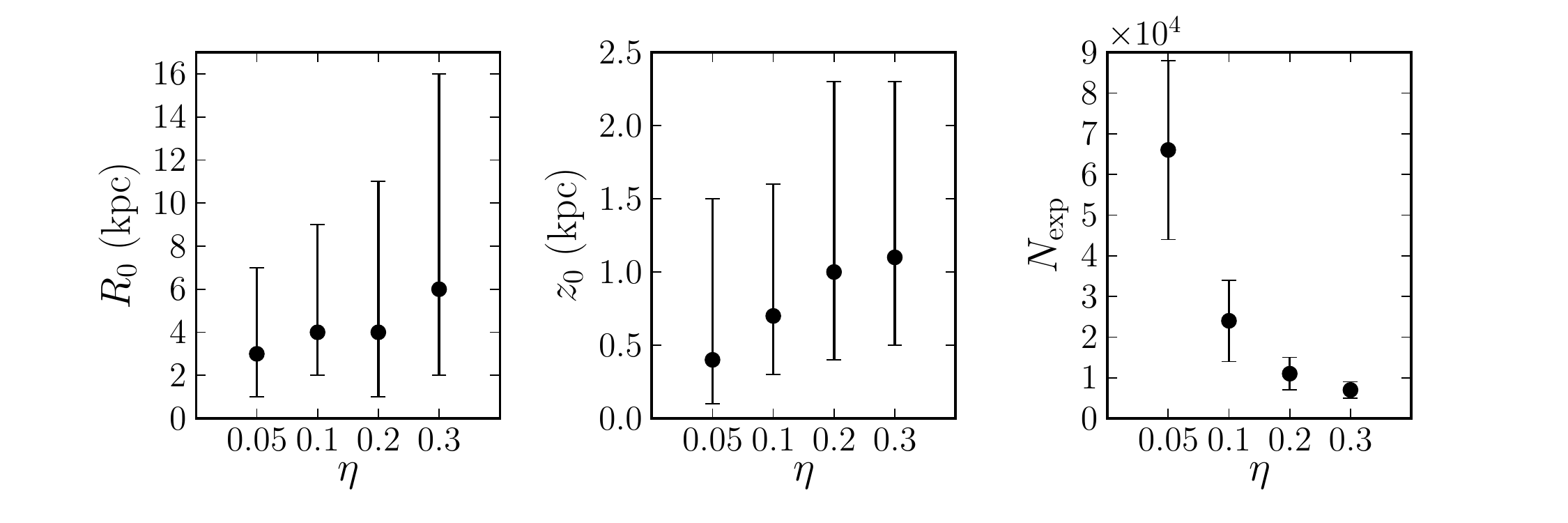}
    \caption{\label{fig:etaeffect}
    Dependecy of $R_{\rm 0}$, $z_{\rm 0}$ and \Nexp\ on the assumed \gam-ray efficiency
    $\eta$ of MSPs (see Eq.~\ref{eq:gammaluminosity}) for model \stexp.
    Error bars are $2\sigma$ confidence.
    }
\end{figure*}

The ratio between flux $F_\mathrm{MSP}$ and sensitivity $F_\mathrm{sensitivity}$
for a given source should approximately scale with the ratio between source
significance ($\sigma$) and chosen significance limit ($\sigma_0$), i.e.
\begin{equation}
\frac{F_\mathrm{MSP}}{F_\mathrm{sensitivity}} \approx \frac{\sigma}{\sigma_0} \approx \sqrt{\frac{TS}{25}}
\end{equation}
For the 1FGL catalogue, source significance is measured using the Test Statistics \citep{Abdo2010},
which roughly scales with the square of the source significance.
The threshold for source detection has been set to $TS=25$ for 1FGL.
We can use these relations to verify how well the actual sensitivity map matches the measured
flux values.
For this purpose, we multiply $F_\mathrm{sensitivity}$ using an arbitrary scaling factor $s$
and determine the value that minimizes the function:
\begin{equation}
\chi^2 = \sum \frac{\left(F_\mathrm{MSP} - s F_\mathrm{sensitivity} \sqrt{TS_\mathrm{MSP}/25} \right)^2}
{\left( \Delta F_\mathrm{MSP} \right)^2},
\end{equation}
where $\Delta F_\mathrm{MSP}$ is the measured statistical uncertainty in the source flux and
$TS_\mathrm{MSP}$ is the Test Statistics of the MSPs.
A best scaling factor of $s \approx 1.4$ results, which is shown in Fig.~\ref{fig:plotSensitivity}
as a dashed line.
Using an 1FGL sensitivity map scaled by a factor of 1.4 reduces the radial scale length 
$R_0$ from $4$ kpc to $3$ kpc and the
vertical scale height 
$z_0$ from $1.0$ kpc to $0.6$ kpc,
while the expected number of \gam-ray emitting MSPs in the Galaxy 
increases from $11\,000$ to $18\,000$.
In general, increasing the sensitivity values 
(i.e. reducing the assumed \fermiLAT\ detection sensitivity)
shuffles more MSPs from the detectable sample into the undetectable sample, 
which leads to larger scale factors $\alpha$ in the maximum likelihood analysis.
Consequently, the
total number \Nexp\ of 
gamma-ray emitting
MSPs in the Galaxy
and the diffuse flux estimates are increased.

For the first \fermiLAT\ pulsar catalogue, \cite{Abdo2010a} presents a sensitivity map
that is specifically computed for sources with pulsar-like spectra and assumes an exponentially cut off
power law with spectral index $\Gamma = 1.4$ and cut off energy $E_{\rm c} = 2.2$~GeV.
We scale this sensitivity from the 6 months of data used by \cite{Abdo2010a} to the 11 months of
data used for 1FGL by dividing the map by $\sqrt{11/6}=1.35$.
Using this sensitivity map, instead of the 1FGL sensitivity map, provides the same radial scale length 
$R_0 = 4$ kpc
and reduces the vertical scale height 
$z_0$ from $1.0$ kpc to $0.8$ kpc,
while the expected number of \gam-ray emitting MSPs in the Galaxy 
increases from $11\,000$ to $16\,000$.

The sensitivity map thus has a non-negligible impact on our analysis, but given the still larger statistical
uncertainties, the precise prescription and scaling of the sensitivity map does not significantly
affect our conclusions.

\subsubsection{Luminosity law\label{sec:LumLaw}}

Another source of uncertainty is related to our specific choice for the luminosity function.
In particular, the precise value $\eta$ for the \gam-ray 
efficiency of MSPs is rather poorly constrained by the present data
as illustrated in Fig.~\ref{fig:FgFedot}.
We thus explore how changes in $\eta$ affect our analysis results
and present the variation of $R_{\rm 0}$, $z_{\rm 0}$ and \Nexp\  for the \stexp\ model for 
efficiencies of $\eta=0.05, 0.1, 0.2,$ and $0.3$ in Fig.~\ref{fig:etaeffect}.

The first trend is that $R_{\rm 0}$ and $z_{\rm 0}$ increase with increasing \gam-ray
efficiency.
We have already explained the reason behind this effect in section \ref{sec:res},
when comparing the results for the {\tt ST} model to those for the {\tt FG} model.
Notably, a larger $\eta$ leads to a larger volume of the Galaxy that becomes detectable
and to a spatial distribution for the detectable MSPs that is more concentrated towards 
the Galactic plane and the central Galactic radian.
To match the observed MSP distribution, the scale parameters thus need to be increased with 
respect to those obtained for lower efficiencies.
The variation over the explored range of $\eta$ amounts to be about a factor of 2, which is
smaller than the statistical uncertainties.

The second trend is that \Nexp\ strongly declines with an increasing value of $\eta$.
This effect also has already been observed in the comparison of the {\tt ST} model and the
{\tt FG} model results and is explained by the increase in the Galactic volume that is
explored by \fermiLAT, which has an increasing $\eta$ for a given sensitivity limit.
Here the effect is significant, that is, the choice of the $\eta$ value dominates the statistical
uncertainties of the MSP sample.
Fortunately, the $\eta$ dependency weakens for an increasing \gam-ray efficiency,
and the change in \Nexp\ becomes rather low above $\ga20\%$.
Figure \ref{fig:FgFedot} illustrates that the average $\eta$ values are found in the range
$0.2-0.3$, and that the values as low as $0.1$ or even $0.05$ appear to be excluded, unless the
distance estimates to MSPs are heavily biased.
Our estimates determined based on the value of $\eta=0.2$ are thus relatively solid and likely constrain the 
true value of \Nexp\ from the high side, since the higher values of $\eta$ are consistent
with the \fermiLAT\ data.


\section{Discussion\label{sec:dis}}

The comparison of the spatial distribution of the \Nmsp\ MSPs detected in the 1FGL catalogue
to a population model provides first constraints on the spatial distribution of Galactic MSPs,
which are based solely on \gam-ray data.
Although the statistical uncertainties still remain large, the analysis suggests radial scale lengths
of $R_0 \sim 4$ kpc for an exponential law and $\sim 6$~kpc for a Gaussian law,
when the modelling approach of \citet{Story2007} is used.
The vertical scale height for this model is around $z_0 \sim 1$ kpc.
The alternative approach of \citet{Faucher-Giguere2010}
suggests higher values; yet, this approach leads to a $P-\dot P$ distribution that seems
incompatible with the \fermiLAT\ data, and that predicts pulsars with characteristic ages
in excess of the age of the Universe.
We thus concentrate in the remainder of this section on the results obtained using the 
{\tt ST} models.

Adding unassociated 1FGL sources with spectral and temporal characteristics reminiscent of
MSPs to the sample reduces the statistical uncertainties, which
suggests an exponential radial scale length of 
$R_0 \sim 3$~kpc 
and an exponential vertical scale height of 
$z_0 \sim 0.6$~kpc.
These values are in the range of expectations for Galactic MSPs \citep{Story2007}.

It should be noted that the estimation of the vertical scale height could be biased 
due to inaccurate modelling of the \fermiLAT\ sensitivity variations close to the Galactic plane.
The sensitivity map only reflects the statistical limitations in the point source detection process,
while inaccurate modelling of the Galactic diffuse emission necessarily adds a systematic uncertainty
that is not quantified.
It is, however, assuring that exclusion of the complex Galactic plane region does not alter our
maximum likelihood results, which demonstrates that our results are robust with respect to analysis 
details.

Another interesting estimate resulting from our analysis is the expected number \Nexp\ of \gam-ray
emitting MSPs in the Galaxy.
It should be emphasized that the total number of MSPs in the Galaxy may indeed be larger than
\Nexp, because a fraction of the MSPs may not be \gam-ray emitters or may have a \gam-ray beam
that does not intersect with the line of sight towards Earth.
Using the more plausible $P-\dot P$ prescription of \citet{Story2007}, our model estimates
\Nexp\ to be $9\,000-11\,000$ with a typical statistical uncertainty of $\pm 4\,000$ (2$\sigma$
confidence).
Althouhg there are more \gam-ray emitting MSPs in the 1FGL catalogue, 
our extended sample of \Nuna\ MSPs raises \Nexp\ to $22\,000 \pm 14\,000$.
Increasing the assumed \gam-ray efficiency $\eta$ above the value of $0.2$ in
this study decreases this number.
Radio estimates of the Galactic MSP population that have been derived by extrapolating the 
local density to the entire Galaxy \citep{Cordes1997, Lyne1998, Lorimer2005},
fall in the interval between $\sim30\,000$ and $\sim200\,000$ objects, which is only marginally 
compatible with our estimates.
Clearly, our \gam-ray estimate based on the \fermiLAT\ detection of MSPs favours lower
numbers.

By assuming a typical lifetime of 10 Gyr for MSPs \citep{Camilo1994}, we can translate our
estimated number of $9\,000 - 11\,000$ Galactic \gam-ray emitting MSPs into a birthrate
of $(0.9-1.1) \times 10^{-6}$ yr$^{-1}$.
This value is 3 times less than the conventional MSP birthrates derived from
radio observations \citep{Cordes1997, Lyne1998, Lorimer2005, Ferrario2007, Story2007}.
On the other hand, birthrates of low-mass X-ray binaries (LMXB), which are suggested to
be the progenitors of MSP \citep{Alpar1982}, are estimated to be $\sim10^{-7}-10^{-6}$ yr$^{-1}$
\citep{Kulkarni1988, Cote1989, Lorimer1995}, and the discrepancy with respect to the estimated
radio MSP birthrates is commonly referred to as the ``birthrate problem''.
Our birthrate estimate is still at the high side of LMXB birthrate estimates, yet marginally 
compatible with the range of proposed values.

\citet{Faucher-Giguere2010} suggested that $\sim10-20\%$ of the high-latitude \gam-ray
background detected by \fermiLAT\ could originate from unresolved MSPs.
Our analysis suggests an average intensity above 100 MeV that amounts to less than
one percent of the high-latitude \gam-ray background emission, and leads us to
conclude that MSPs do probably not provide any significant contribution to the isotropic diffuse
\gam-ray background.

In contrast, a few percent of the Galactic diffuse emission towards the inner central radian may
be attributed to unresolved MSPs, although it remains questionable whether such a component 
can reliably be extracted from the dominating cosmic-ray induced components.
Possibly, the characteristic spectral shape (an exponentially cut-off power law with a hard spectral
index of $\Gamma \sim 1.4$ and a cut-off energy of a few GeV) of MSP could be used to disentangle
the components.
Additional help may come from the relative large vertical scale height of MSPs, which exceeds
those of the gas related cosmic-ray induced components (pion decay, Bremsstrahlung).
Conversely, the Inverse Compton component is expected to reach towards higher latitudes,
which makes a spatial distinction between MSP and cosmic-ray induced \gam-ray emission
challenging.

\section{Conclusion \label{sec:conclusion}}

Based on a Monte Carlo model, we derive constraints on the Galactic MSP
population based solely on \gam-ray detections of MSPs for the first time.
We use a maximum likelihood method to determine the radial scale length, vertical
scale height, and total size of the Galactic \gam-ray emitting MSP population.
We apply our model to a sample of \Nmsp\ MSPs that are detected in the 1FGL catalogue,
and to an increased list of \Nuna\ objects comprised of confirmed MSPs and MSP candidates.

The spatial parameters are consistent with previous estimates based on radio observations.
The estimated size of the Galactic MSP population is lower than the estimates obtained
from radio observations, which alleviates the MSP birthrate problem.
The contribution of unresolved MSPs to the isotropic diffuse background emission at high
Galactic latitudes is estimated to be negligible.
Towards the central Galactic radian, unresolved MSPs may contribute up to a few percent
of the total emission. 
It remains to be seen, however, whether this contribution can be reliably disentangled from 
the other dominating emission components.

Our analysis still suffers from relatively large statistical uncertainties, because of the limited number
of MSPs detected so far in \gam-rays.
Using an extended sample that also includes unassociated 1FGL catalogue sources with
spectral and temporal characteristics reminiscent of MSPs leads
to a substantial reduction of the uncertainties.
Assuming that the number of detected MSPs by \fermiLAT\ simply scales with the decrease
of sensitivity that arises from a continuous survey of the sky, we expect that
$\sim80$ MSPs should be detected by \fermiLAT\ after 5 years of observations.
This number should increase to $\sim110$ MSPs after 10 years.
As shown from the analysis of our extended sample, such a large number of \gam-ray
detections will provide us with unprecedented constraints on the Galactic MSP
distribution.

\begin{acknowledgements}
  Some of the results in this paper have been derived using the HEALPix (K.M.~G\'orski et al., 2005, ApJ, 622, p759) package.
\end{acknowledgements}

\bibliography{biblio}

\begin{thebibliography}{41}
\expandafter\ifx\csname natexlab\endcsname\relax\def\natexlab#1{#1}\fi

\bibitem[{Abdo {et~al.}(2010{\natexlab{a}})Abdo, Ackermann, Ajello, Allafort,
  Antolini, Atwood, Axelsson, Baldini, Ballet, Barbiellini, Bastieri, Baughman,
  Bechtol, Bellazzini, Belli, Berenji, Bisello, Blandford, Bloom, Bonamente,
  Bonnell, Borgland, Bouvier, Bregeon, Brez, Brigida, Bruel, Burnett, Busetto,
  Buson, Caliandro, Cameron, Campana, Canadas, Caraveo, Carrigan, Casandjian,
  Cavazzuti, Ceccanti, Cecchi, \c{C}elik, Charles, Chekhtman, Cheung, Chiang,
  Cillis, Ciprini, Claus, Cohen-Tanugi, Conrad, Corbet, Davis, DeKlotz, den
  Hartog, Dermer, de~Angelis, de~Luca, de~Palma, Digel, Dormody, Silva, Drell,
  Dubois, Dumora, Fabiani, Farnier, Favuzzi, Fegan, Ferrara, Focke, Fortin,
  Frailis, Fukazawa, Funk, Fusco, Gargano, Gasparrini, Gehrels, Germani,
  Giavitto, Giebels, Giglietto, Giommi, Giordano, Giroletti, Glanzman, Godfrey,
  Grenier, Grondin, Grove, Guillemot, Guiriec, Gustafsson, Hadasch, Hanabata,
  Harding, Hayashida, Hays, Healey, Hill, Horan, Hughes, Iafrate,
  J\'ohannesson, Johnson, Johnson, Johnson, Johnson, Kamae, Katagiri, Kataoka,
  Kawai, Kerr, Kn\"odlseder, Kocevski, Kuss, Lande, Landriu, Latronico, Lee,
  Lemoine-Goumard, Lionetto, Llena~Garde, Longo, Loparco, Lott, Lovellette,
  Lubrano, Madejski, Makeev, Marangelli, Marelli, Massaro, Mazziotta,
  McConville, McEnery, Michelson, Minuti, Mitthumsiri, Mizuno, Moiseev,
  Mongelli, Monte, Monzani, Moretti, Morselli, Moskalenko, Murgia, Nakajima,
  Nakamori, Naumann-Godo, Nolan, Norris, Nuss, Ohno, Ohsugi, Omodei, Orlando,
  Ormes, Ozaki, Paccagnella, Paneque, Panetta, Parent, Pelassa, Pepe,
  Pesce-Rollins, Pinchera, Piron, Porter, Poupard, Rain\`o, Rando, Ray,
  Razzano, Razzaque, Rea, Reimer, Reimer, Reposeur, Ripken, Ritz, Rochester,
  Rodriguez, Romani, Roth, Sadrozinski, Salvetti, Sanchez, Sander,
  Saz~Parkinson, Scargle, Schalk, Scolieri, Sgr\`o, Shaw, Siskind, Smith,
  Smith, Spandre, Spinelli, Starck, Stephens, Striani, Strickman, Strong,
  Suson, Tajima, Takahashi, Takahashi, Tanaka, Thayer, Thayer, Thompson,
  Tibaldo, Tibolla, Tinebra, Torres, Tosti, Tramacere, Uchiyama, Usher,
  Van~Etten, Vasileiou, Vilchez, Vitale, Waite, Wallace, Wang, Watters, Winer,
  Wood, Yang, Ylinen, \& Ziegler}]{Abdo2010}
Abdo, A.~A., Ackermann, M., Ajello, M., {et~al.} 2010{\natexlab{a}}, The
  Astrophysical Journal Supplement Series, 188, 405

\bibitem[{Abdo {et~al.}(2010{\natexlab{b}})Abdo, {Ackermann}, {Ajello},
  {Allafort}, {Baldini}, {Ballet}, {Barbiellini}, {Bastieri}, {Bechtol},
  {Bellazzini}, {Berenji}, {Blandford}, {Bloom}, {Bonamente}, {Borgland},
  {Bouvier}, {Bregeon}, {Brez}, {Brigida}, {Bruel}, {Burnett}, {Buson},
  {Caliandro}, {Cameron}, {Camilo}, {Caraveo}, {Carrigan}, {Casandjian},
  {Cecchi}, {{\c C}elik}, {Chekhtman}, {Cheung}, {Chiang}, {Ciprini}, {Claus},
  {Cognard}, {Cohen-Tanugi}, {Conrad}, {Corbet}, {DeCesar}, {Dermer},
  {Desvignes}, {de Angelis}, {de Palma}, {Digel}, {Dormody}, {Silva}, {Drell},
  {Dubois}, {Dumora}, {Espinoza}, {Farnier}, {Favuzzi}, {Fegan}, {Focke},
  {Frailis}, {Freire}, {Fukazawa}, {Funk}, {Fusco}, {Gargano}, {Gasparrini},
  {Gehrels}, {Germani}, {Giavitto}, {Giglietto}, {Giordano}, {Glanzman},
  {Godfrey}, {Grenier}, {Grondin}, {Grove}, {Guillemot}, {Guiriec}, {Hadasch},
  {Harding}, {Hays}, {Hobbs}, {Horan}, {Hughes}, {J{\'o}hannesson}, {Johnson},
  {Johnson}, {Johnson}, {Johnston}, {Kamae}, {Katagiri}, {Kataoka}, {Kawai},
  {Kerr}, {Kn{\"o}dlseder}, {Kramer}, {Kuss}, {Lande}, {Latronico},
  {Lemoine-Goumard}, {Llena Garde}, {Longo}, {Loparco}, {Lott}, {Lovellette},
  {Lubrano}, {Lyne}, {Makeev}, {Manchester}, {Marelli}, {Mazziotta},
  {McConville}, {McEnery}, {McGlynn}, {Meurer}, {Michelson}, {Mitthumsiri},
  {Mizuno}, {Moiseev}, {Monte}, {Monzani}, {Morselli}, {Moskalenko}, {Murgia},
  {Nolan}, {Norris}, {Noutsos}, {Nuss}, {Ohsugi}, {Omodei}, {Orlando}, {Ormes},
  {Ozaki}, {Paneque}, {Panetta}, {Parent}, {Pelassa}, {Pepe}, {Pesce-Rollins},
  {Pierbattista}, {Piron}, {Porter}, {Rain{\`o}}, {Rando}, {Ransom}, {Razzano},
  {Reimer}, {Reimer}, {Reposeur}, {Ripken}, {Ritz}, {Rochester}, {Rodriguez},
  {Romani}, {Roth}, {Ryde}, {Sadrozinski}, {Sander}, {Saz Parkinson},
  {Scargle}, {Sgr{\`o}}, {Siskind}, {Smith}, {Smith}, {Spandre}, {Spinelli},
  {Stappers}, {Starck}, {Strickman}, {Suson}, {Takahashi}, {Tanaka}, {Thayer},
  {Thayer}, {Theureau}, {Thompson}, {Thorsett}, {Tibaldo}, {Torres}, {Tosti},
  {Tramacere}, {Usher}, {Van Etten}, {Vasileiou}, {Venter}, {Vilchez},
  {Vitale}, {Waite}, {Wallace}, {Wang}, {Weltevrede}, {Winer}, {Wood},
  {Ylinen}, \& {Ziegler}}]{Abdo2010c}
Abdo, A.~A., {Ackermann}, M., {Ajello}, M., {et~al.} 2010{\natexlab{b}}, The
  Astrophysical Journal, 712, 957

\bibitem[{Abdo {et~al.}(2009{\natexlab{a}})Abdo, Ackermann, Ajello, Atwood,
  Axelsson, Baldini, Ballet, Barbiellini, Baring, Bastieri, Baughman, Bechtol,
  Bellazzini, Berenji, Bignami, Blandford, Bloom, Bonamente, Borgland, Bregeon,
  Brez, Brigida, Bruel, Burnett, Caliandro, Cameron, Camilo, Caraveo, Carlson,
  Casandjian, Cecchi, Celik, Charles, Chekhtman, Cheung, Chiang, Ciprini,
  Claus, Cognard, Cohen-Tanugi, Cominsky, Conrad, Corbet, Cutini, Dermer,
  Desvignes, de~Angelis, de~Luca, de~Palma, Digel, Dormody, {do Couto e Silva},
  Drell, Dubois, Dumora, Edmonds, Farnier, Favuzzi, Fegan, Focke, Frailis,
  Freire, Fukazawa, Funk, Fusco, Gargano, Gasparrini, Gehrels, Germani,
  Giebels, Giglietto, Giordano, Glanzman, Godfrey, Grenier, Grondin, Grove,
  Guillemot, Guiriec, Hanabata, Harding, Hayashida, Hays, Hobbs, Hughes,
  J\'{o}hannesson, Johnson, Johnson, Johnson, Johnson, Johnston, Kamae,
  Katagiri, Kataoka, Kawai, Kerr, Kn\"{o}dlseder, Kocian, Kramer, Kuss, Lande,
  Latronico, Lemoine-Goumard, Longo, Loparco, Lott, Lovellette, Lubrano,
  Madejski, Makeev, Manchester, Marelli, Mazziotta, McConville, McEnery,
  McLaughlin, Meurer, Michelson, Mitthumsiri, Mizuno, Moiseev, Monte, Monzani,
  Morselli, Moskalenko, Murgia, Nolan, Norris, Nuss, Ohsugi, Omodei, Orlando,
  Ormes, Paneque, Panetta, Parent, Pelassa, Pepe, Pesce-Rollins, Piron, Porter,
  Rain\`{o}, Rando, Ransom, Ray, Razzano, Rea, Reimer, Reimer, Reposeur, Ritz,
  Rochester, Rodriguez, Romani, Roth, Ryde, Sadrozinski, Sanchez, Sander, {Saz
  Parkinson}, Scargle, Schalk, Sgr\`{o}, Siskind, Smith, Smith, Spandre,
  Spinelli, Stappers, Starck, Striani, Strickman, Suson, Tajima, Takahashi,
  Tanaka, Thayer, Thayer, Theureau, Thompson, Thorsett, Tibaldo, Torres, Tosti,
  Tramacere, Uchiyama, Usher, {Van Etten}, Vasileiou, Venter, Vilchez, Vitale,
  Waite, Wallace, Wang, Watters, Webb, Weltevrede, Winer, Wood, Ylinen, \&
  Ziegler}]{Abdo2009}
Abdo, a.~a., Ackermann, M., Ajello, M., {et~al.} 2009{\natexlab{a}}, Science
  (New York, N.Y.), 325, 848

\bibitem[{Abdo {et~al.}(2010{\natexlab{c}})Abdo, Ackermann, Ajello, Atwood,
  Axelsson, Baldini, Ballet, Barbiellini, Baring, Bastieri, Baughman, Bechtol,
  Bellazzini, Berenji, Blandford, Bloom, Bonamente, Borgland, Bregeon, Brez,
  Brigida, Bruel, Burnett, Buson, Caliandro, Cameron, Camilo, Caraveo,
  Casandjian, Cecchi, Charles, Chekhtman, Cheung, Chiang, Ciprini, Claus,
  Cognard, Cominsky, Conrad, Corbet, Cutini, Hartog, Dermer, Angelis, Luca,
  Palma, Digel, Dormody, Drell, Dubois, Dumora, Espinoza, Farnier, Favuzzi,
  Fegan, Ferrara, Focke, Fortin, Frailis, Freire, Fukazawa, Funk, Fusco,
  Gargano, Gasparrini, Gehrels, Germani, Giavitto, Giebels, Giglietto, Giommi,
  Giordano, Glanzman, Godfrey, Gotthelf, Grenier, Grove, Guillemot, Guiriec,
  Gwon, Hanabata, Harding, Hayashida, Hays, Hughes, Jackson, Johnson, Johnson,
  Johnson, Johnson, Johnston, Kamae, Kanbach, Kaspi, Katagiri, Kataoka, Kawai,
  Kerr, Kocian, Kramer, Kuss, Lande, Latronico, Livingstone, Longo, Loparco,
  Lott, Lovellette, Lubrano, Lyne, Madejski, Makeev, Manchester, Marelli,
  Mazziotta, Mcconville, Mcenery, Mcglynn, Meurer, Michelson, Mineo,
  Mitthumsiri, Mizuno, Moiseev, Monte, Monzani, Morselli, Moskalenko, Murgia,
  Nakamori, Nolan, Norris, Noutsos, Nuss, Ohsugi, Omodei, Orlando, Ormes,
  Ozaki, Paneque, Panetta, Parent, Pelassa, Pepe, Piron, Porter, Rando, Ransom,
  Ray, Razzano, Rea, Reimer, Reimer, Reposeur, Ritz, Rodriguez, Romani, Roth,
  Ryde, Sanchez, Sander, Parkinson, Scargle, Schalk, Sellerholm, Siskind,
  Smith, Smith, Spandre, Spinelli, Stappers, Striani, Strickman, Strong, Suson,
  Tajima, Takahashi, Takahashi, Tanaka, Thayer, Thayer, Theureau, Thompson,
  Thorsett, Tibaldo, Tibolla, Torres, Tosti, Tramacere, Uchiyama, Usher, Etten,
  Vasileiou, Venter, Vilchez, Vitale, Waite, Wang, Wang, Watters, Weltevrede,
  Winer, Wood, Ylinen, \& Ziegler}]{Abdo2010a}
Abdo, A.~A., Ackermann, M., Ajello, M., {et~al.} 2010{\natexlab{c}}, The
  Astrophysical Journal Supplement

\bibitem[{{Abdo} {et~al.}(2010){Abdo}, {Ackermann}, {Ajello}, {Atwood},
  {Baldini}, {Ballet}, {Barbiellini}, {Bastieri}, {Baughman}, {Bechtol},
  {Bellazzini}, {Berenji}, {Blandford}, {Bloom}, {Bonamente}, {Borgland},
  {Bregeon}, {Brez}, {Brigida}, {Bruel}, {Burnett}, {Buson}, {Caliandro},
  {Cameron}, {Caraveo}, {Casandjian}, {Cavazzuti}, {Cecchi}, {{\c C}elik},
  {Charles}, {Chekhtman}, {Cheung}, {Chiang}, {Ciprini}, {Claus},
  {Cohen-Tanugi}, {Cominsky}, {Conrad}, {Cutini}, {Dermer}, {de Angelis}, {de
  Palma}, {Digel}, {di Bernardo}, {do Couto e Silva}, {Drell}, {Drlica-Wagner},
  {Dubois}, {Dumora}, {Farnier}, {Favuzzi}, {Fegan}, {Focke}, {Fortin},
  {Frailis}, {Fukazawa}, {Funk}, {Fusco}, {Gaggero}, {Gargano}, {Gasparrini},
  {Gehrels}, {Germani}, {Giebels}, {Giglietto}, {Giommi}, {Giordano},
  {Glanzman}, {Godfrey}, {Grenier}, {Grondin}, {Grove}, {Guillemot}, {Guiriec},
  {Gustafsson}, {Hanabata}, {Harding}, {Hayashida}, {Hughes}, {Itoh},
  {Jackson}, {J{\'o}hannesson}, {Johnson}, {Johnson}, {Johnson}, {Johnson},
  {Kamae}, {Katagiri}, {Kataoka}, {Kawai}, {Kerr}, {Kn{\"o}dlseder}, {Kocian},
  {Kuehn}, {Kuss}, {Lande}, {Latronico}, {Lemoine-Goumard}, {Longo}, {Loparco},
  {Lott}, {Lovellette}, {Lubrano}, {Madejski}, {Makeev}, {Mazziotta},
  {McConville}, {McEnery}, {Meurer}, {Michelson}, {Mitthumsiri}, {Mizuno},
  {Moiseev}, {Monte}, {Monzani}, {Morselli}, {Moskalenko}, {Murgia}, {Nolan},
  {Norris}, {Nuss}, {Ohsugi}, {Omodei}, {Orlando}, {Ormes}, {Paneque},
  {Panetta}, {Parent}, {Pelassa}, {Pepe}, {Pesce-Rollins}, {Piron}, {Porter},
  {Rain{\`o}}, {Rando}, {Razzano}, {Reimer}, {Reimer}, {Reposeur}, {Ritz},
  {Rochester}, {Rodriguez}, {Roth}, {Ryde}, {Sadrozinski}, {Sanchez}, {Sander},
  {Parkinson}, {Scargle}, {Sellerholm}, {Sgr{\`o}}, {Shaw}, {Siskind}, {Smith},
  {Smith}, {Spandre}, {Spinelli}, {Starck}, {Strickman}, {Strong}, {Suson},
  {Tajima}, {Takahashi}, {Takahashi}, {Tanaka}, {Thayer}, {Thayer}, {Thompson},
  {Tibaldo}, {Torres}, {Tosti}, {Tramacere}, {Uchiyama}, {Usher}, {Vasileiou},
  {Vilchez}, {Vitale}, {Waite}, {Wang}, {Winer}, {Wood}, {Ylinen}, {Ziegler},
  \& {Fermi LAT Collaboration}}]{Abdo2010EB}
{Abdo}, A.~A., {Ackermann}, M., {Ajello}, M., {et~al.} 2010, Physical Review
  Letters, 104, 101101

\bibitem[{Abdo {et~al.}(2009{\natexlab{b}})Abdo, Ackermann, Atwood, Axelsson,
  Baldini, Ballet, Barbiellini, Bastieri, Battelino, Baughman, Bechtol,
  Bellazzini, Berenji, Bloom, Bonamente, Borgland, Bregeon, Brez, Brigida,
  Bruel, Burnett, Caliandro, Cameron, Caraveo, Casandjian, Cecchi, Charles,
  Chekhtman, Cheung, Chiang, Ciprini, Claus, Cognard, Cohen-Tanugi, Cominsky,
  Conrad, Cutini, Dermer, de~Angelis, de~Palma, Digel, Dormody, do~Couto~e
  Silva, Drell, Dubois, Dumora, Farnier, Favuzzi, Focke, Frailis, Fukazawa,
  Funk, Fusco, Gargano, Gasparrini, Gehrels, Germani, Giebels, Giglietto,
  Giordano, Glanzman, Godfrey, Grenier, Grondin, Grove, Guillemot, Guiriec,
  Hanabata, Harding, Hayashida, Hays, Hughes, J{\'o}hannesson, Johnson,
  Johnson, Johnson, Johnson, Kamae, Katagiri, Kataoka, Kawai, Kerr,
  Kn{\"o}dlseder, Kocian, Komin, Kuehn, Kuss, Lande, Latronico, Lee,
  Lemoine-Goumard, Longo, Loparco, Lott, Lovellette, Lubrano, Madejski, Makeev,
  Marelli, Mazziotta, McConville, McEnery, Meurer, Michelson, Mitthumsiri,
  Mizuno, Moiseev, Monte, Monzani, Morselli, Moskalenko, Murgia, Nolan, Nuss,
  Ohsugi, Omodei, Orlando, Ormes, Pancrazi, Paneque, Panetta, Parent, Pepe,
  Pesce-Rollins, Piron, Porter, Rain{\`o}, Rando, Razzano, Reimer, Reimer,
  Reposeur, Ritz, Rochester, Rodriguez, Romani, Ryde, Sadrozinski, Sanchez,
  Sander, Parkinson, Sgr{\`o}, Siskind, Smith, Smith, Spandre, Spinelli,
  Starck, Strickman, Suson, Tajima, Takahashi, Tanaka, Thayer, Thayer,
  Theureau, Thompson, Tibaldo, Torres, Tosti, Tramacere, Uchiyama, Usher,
  Etten, Vilchez, Vitale, Waite, Watters, Webb, Wood, Ylinen, \&
  Ziegler}]{Abdo2009c}
Abdo, A.~A., Ackermann, M., Atwood, W.~B., {et~al.} 2009{\natexlab{b}}, The
  Astrophysical Journal, 699, 1171

\bibitem[{{Ackermann} {et~al.}(2012){Ackermann}, {Ajello}, {Atwood}, {Baldini},
  {Ballet}, {Barbiellini}, {Bastieri}, {Bechtol}, {Bellazzini}, {Berenji},
  {Blandford}, {Bloom}, {Bonamente}, {Borgland}, {Brandt}, {Bregeon},
  {Brigida}, {Bruel}, {Buehler}, {Buson}, {Caliandro}, {Cameron}, {Caraveo},
  {Cavazzuti}, {Cecchi}, {Charles}, {Chekhtman}, {Chiang}, {Ciprini}, {Claus},
  {Cohen-Tanugi}, {Conrad}, {Cutini}, {de Angelis}, {de Palma}, {Dermer},
  {Digel}, {Silva}, {Drell}, {Drlica-Wagner}, {Falletti}, {Favuzzi}, {Fegan},
  {Ferrara}, {Focke}, {Fortin}, {Fukazawa}, {Funk}, {Fusco}, {Gaggero},
  {Gargano}, {Germani}, {Giglietto}, {Giordano}, {Giroletti}, {Glanzman},
  {Godfrey}, {Grove}, {Guiriec}, {Gustafsson}, {Hadasch}, {Hanabata},
  {Harding}, {Hayashida}, {Hays}, {Horan}, {Hou}, {Hughes}, {J{\'o}hannesson},
  {Johnson}, {Johnson}, {Kamae}, {Katagiri}, {Kataoka}, {Kn{\"o}dlseder},
  {Kuss}, {Lande}, {Latronico}, {Lee}, {Lemoine-Goumard}, {Longo}, {Loparco},
  {Lott}, {Lovellette}, {Lubrano}, {Mazziotta}, {McEnery}, {Michelson},
  {Mitthumsiri}, {Mizuno}, {Monte}, {Monzani}, {Morselli}, {Moskalenko},
  {Murgia}, {Naumann-Godo}, {Norris}, {Nuss}, {Ohsugi}, {Okumura}, {Omodei},
  {Orlando}, {Ormes}, {Paneque}, {Panetta}, {Parent}, {Pesce-Rollins},
  {Pierbattista}, {Piron}, {Pivato}, {Porter}, {Rain{\`o}}, {Rando}, {Razzano},
  {Razzaque}, {Reimer}, {Reimer}, {Sadrozinski}, {Sgr{\`o}}, {Siskind},
  {Spandre}, {Spinelli}, {Strong}, {Suson}, {Takahashi}, {Tanaka}, {Thayer},
  {Thayer}, {Thompson}, {Tibaldo}, {Tinivella}, {Torres}, {Tosti}, {Troja},
  {Usher}, {Vandenbroucke}, {Vasileiou}, {Vianello}, {Vitale}, {Waite}, {Wang},
  {Winer}, {Wood}, {Wood}, {Yang}, {Ziegler}, \& {Zimmer}}]{Ackermann2012a}
{Ackermann}, M., {Ajello}, M., {Atwood}, W.~B., {et~al.} 2012, The
  Astrophysical Journal, 750, 3

\bibitem[{Alpar {et~al.}(1982)Alpar, Cheng, Ruderman, \& Shaham}]{Alpar1982}
Alpar, M.~A., Cheng, A.~F., Ruderman, M.~A., \& Shaham, J. 1982, Nature, 300,
  728

\bibitem[{Bhattacharya \& Srinivasan(1991)}]{Bhattacharya1991}
Bhattacharya, D. \& Srinivasan, G. 1991, Journal of Astrophysics and Astronomy,
  12, 17

\bibitem[{{Camilo} {et~al.}(1994){Camilo}, {Thorsett}, \&
  {Kulkarni}}]{Camilo1994}
{Camilo}, F., {Thorsett}, S.~E., \& {Kulkarni}, S.~R. 1994, \apjl, 421, L15

\bibitem[{Cash(1979)}]{Cash1979}
Cash, W. 1979, The Astrophysical Journal, 228, 939

\bibitem[{Cognard {et~al.}(2011)Cognard, Guillemot, Johnson, Smith, Venter,
  Harding, Wolff, Cheung, Donato, Abdo, Ballet, Camilo, Desvignes, Dumora,
  Ferrara, Freire, Grove, Johnston, Keith, Kramer, Lyne, Michelson, Parent,
  Ransom, Ray, Romani, Parkinson, Stappers, Theureau, Thompson, Weltevrede, \&
  Wood}]{Cognard2011}
Cognard, I., Guillemot, L., Johnson, T.~J., {et~al.} 2011, The Astrophysical
  Journal, 732, 47

\bibitem[{Cordes \& Chernoff(1997)}]{Cordes1997}
Cordes, J.~M. \& Chernoff, D.~F. 1997, The Astrophysical Journal, 482

\bibitem[{{Cote} \& {Pylyser}(1989)}]{Cote1989}
{Cote}, J. \& {Pylyser}, E.~H.~P. 1989, \aap, 218, 131

\bibitem[{Efron(1979)}]{Efron1979}
Efron, B. 1979, The Annals of Statistics, 7, 1

\bibitem[{Faucher-Gigu\`{e}re \& Loeb(2010)}]{Faucher-Giguere2010}
Faucher-Gigu\`{e}re, C.-A. \& Loeb, A. 2010, Journal of Cosmology and
  Astroparticle Physics, 2010, 005

\bibitem[{{Ferrario} \& {Wickramasinghe}(2007)}]{Ferrario2007}
{Ferrario}, L. \& {Wickramasinghe}, D. 2007, \mnras, 375, 1009

\bibitem[{G{\'o}rski {et~al.}(2005)G{\'o}rski, Hivon, Banday, Wandelt, Hansen,
  Reinecke, \& Bartelmann}]{Gorski2005}
G{\'o}rski, K.~M., Hivon, E., Banday, A.~J., {et~al.} 2005, The Astrophysical
  Journal, 622, 759

\bibitem[{Guillemot(2010)}]{Guillemot2010}
Guillemot, L. 2010, in Radio Pulsars: An astrophysical key to unlock the secret
  of the Universe

\bibitem[{Guillemot {et~al.}(2012{\natexlab{a}})Guillemot, Freire, Cognard,
  Johnson, Takahashi, Kataoka, Desvignes, Camilo, Ferrara, Harding, Janssen,
  Keith, Kerr, Kramer, Parent, Ransom, Ray, Saz~Parkinson, Smith, Stappers, \&
  Theureau}]{Guillemot2012}
Guillemot, L., Freire, P. C.~C., Cognard, I., {et~al.} 2012{\natexlab{a}},
  Monthly Notices of the Royal Astronomical Society, no

\bibitem[{Guillemot {et~al.}(2012{\natexlab{b}})Guillemot, J.~Johnson, Venter,
  Kerr, Pancrazi, Livingstone, H.~Janssen, Jaroenjittichai, Kramer, Cognard,
  W.~Stappers, K.~Harding, Camilo, M.~Espinoza, C.~C.~Freire, Gargano,
  E.~Grove, Johnston, F.~Michelson, Noutsos, Parent, M.~Ransom, S.~Ray,
  Shannon, A.~Smith, Theureau, E.~Thorsett, \& Webb}]{Guillemot2012a}
Guillemot, L., J.~Johnson, T., Venter, C., {et~al.} 2012{\natexlab{b}}, The
  Astrophysical Journal, 744, 33

\bibitem[{Hessels {et~al.}(2010)Hessels, Roberts, McLaughlin, Ray, Bangale,
  Ransom, Kerr, Camilo, DeCesar, \& PSC}]{Hessels2010}
Hessels, J. W.~T., Roberts, M. S.~E., McLaughlin, M.~A., {et~al.} 2010, in
  Radio Pulsars: An astrophysical key to unlock the secret of the Universe

\bibitem[{{Hewish} {et~al.}(1968){Hewish}, {Bell}, {Pilkington}, {Scott}, \&
  {Collins}}]{Hewish1968}
{Hewish}, A., {Bell}, S.~J., {Pilkington}, J.~D.~H., {Scott}, P.~F., \&
  {Collins}, R.~A. 1968, Nature, 217, 709

\bibitem[{Keith {et~al.}(2012)Keith, Johnston, Bailes, Bates, Bhat, Burgay,
  Burke-Spolaor, D'Amico, Jameson, Kramer, Levin, Milia, Possenti, Stappers,
  van Straten, \& Parent}]{Keith2012}
Keith, M.~J., Johnston, S., Bailes, M., {et~al.} 2012, Monthly Notices of the
  Royal Astronomical Society, 419, 1752

\bibitem[{Keith {et~al.}(2011)Keith, Johnston, Ray, Ferrara, Saz~Parkinson,
  \c~Celik, Belfiore, Donato, Cheung, Abdo, Camilo, Freire, Guillemot, Harding,
  Kramer, Michelson, Ransom, Romani, Smith, Thompson, Weltevrede, \&
  Wood}]{Keith2011}
Keith, M.~J., Johnston, S., Ray, P.~S., {et~al.} 2011, Monthly Notices of the
  Royal Astronomical Society, 414, 1292

\bibitem[{Kerr {et~al.}(2012)Kerr, Camilo, Johnson, Ferrara, Guillemot,
  Harding, Hessels, Johnston, Keith, Kramer, Ransom, Ray, Reynolds, Sarkissian,
  \& Wood}]{Kerr2012}
Kerr, M., Camilo, F., Johnson, T.~J., {et~al.} 2012, The Astrophysical Journal
  Letters, 748, L2

\bibitem[{{Kuiper} {et~al.}(2000){Kuiper}, {Hermsen}, {Verbunt}, {Thompson},
  {Stairs}, {Lyne}, {Strickman}, \& {Cusumano}}]{Kuiper2000}
{Kuiper}, L., {Hermsen}, W., {Verbunt}, F., {et~al.} 2000, Astronomy \&
  Astrophysics, 359, 615

\bibitem[{{Kulkarni} \& {Narayan}(1988)}]{Kulkarni1988}
{Kulkarni}, S.~R. \& {Narayan}, R. 1988, \apj, 335, 755

\bibitem[{{Lorimer}(1995)}]{Lorimer1995}
{Lorimer}, D.~R. 1995, \mnras, 274, 300

\bibitem[{{Lorimer}(2005)}]{Lorimer2005}
{Lorimer}, D.~R. 2005, Living Reviews in Relativity, 8, 7

\bibitem[{Lyne(2000)}]{Lyne2000}
Lyne, a.~G. 2000, Philosophical Transactions of the Royal Society A:
  Mathematical, Physical and Engineering Sciences, 358, 831

\bibitem[{{Lyne} {et~al.}(1998){Lyne}, {Manchester}, {Lorimer}, {Bailes},
  {D'Amico}, {Tauris}, {Johnston}, {Bell}, \& {Nicastro}}]{Lyne1998}
{Lyne}, A.~G., {Manchester}, R.~N., {Lorimer}, D.~R., {et~al.} 1998, \mnras,
  295, 743

\bibitem[{{Manchester} {et~al.}(2005){Manchester}, {Hobbs}, {Teoh}, \&
  {Hobbs}}]{ATNFpaper}
{Manchester}, R.~N., {Hobbs}, G.~B., {Teoh}, A., \& {Hobbs}, M. 2005, The
  Astronomical Journal, 129, 1993

\bibitem[{Parent(2010)}]{Parent2010}
Parent, D. 2010, in Radio Pulsars: An astrophysical key to unlock the secret of
  the Universe

\bibitem[{{Ransom} {et~al.}(2011){Ransom}, {Ray}, {Camilo}, {Roberts}, {{\c
  C}elik}, {Wolff}, {Cheung}, {Kerr}, {Pennucci}, {DeCesar}, {Cognard}, {Lyne},
  {Stappers}, {Freire}, {Grove}, {Abdo}, {Desvignes}, {Donato}, {Ferrara},
  {Gehrels}, {Guillemot}, {Gwon}, {Harding}, {Johnston}, {Keith}, {Kramer},
  {Michelson}, {Parent}, {Saz Parkinson}, {Romani}, {Smith}, {Theureau},
  {Thompson}, {Weltevrede}, {Wood}, \& {Ziegler}}]{Ransom2011}
{Ransom}, S.~M., {Ray}, P.~S., {Camilo}, F., {et~al.} 2011, The Astrophysical
  Journal Letters, 727

\bibitem[{Ray {et~al.}(2011)Ray, Abdo, Parent, Bhattacharya, Bhattacharyya,
  Camilo, Cognard, Theureau, Harding, Thompson, Freire, Guillemot, Gupta, Roy,
  Hessels, Keith, Shannon, Michelson, Romani, Kramer, McLaughlin, Ransom,
  Roberts, Parkinson, Ziegler, Smith, Stappers, Weltevrede, \& Wood}]{Ray2011a}
Ray, P.~S., Abdo, A.~A., Parent, D., {et~al.} 2011, in The Fermi Symposium

\bibitem[{Reed(2006)}]{Reed2006}
Reed, B.~C. 2006, The Royal Astronomical Society of Canada, 1

\bibitem[{{Robin} {et~al.}(2003){Robin}, {Reyl{\'e}}, {Derri{\`e}re}, \&
  {Picaud}}]{Robin2003}
{Robin}, A.~C., {Reyl{\'e}}, C., {Derri{\`e}re}, S., \& {Picaud}, S. 2003,
  \aap, 409, 523

\bibitem[{Story {et~al.}(2007)Story, Gonthier, \& Harding}]{Story2007}
Story, S.~A., Gonthier, P.~L., \& Harding, A.~K. 2007, The Astrophysical
  Journal, 671, 713

\bibitem[{Teague(1980)}]{Teague1980}
Teague, M.~R. 1980, Journal of the Optical Society of America, 70, 920

\bibitem[{Wilks(1938)}]{Wilks1938}
Wilks, S.~S. 1938, The Annals of Mathematical Statistics, 9, 60

\end{thebibliography}

\end{document}